\documentclass[draftclsnofoot]{IEEEtran}
 \onecolumn
\usepackage{latexsym}
\usepackage{cite}
\usepackage{color}
\usepackage{bm}
\usepackage{amsmath, amssymb}
\usepackage[dvips]{graphics}
\usepackage{graphicx}
\usepackage{psfrag}
 \allowdisplaybreaks

\newtheorem{definition}{Definition}

\newtheorem{theorem}{Theorem}
\newtheorem{lemma}[theorem]{Lemma}

\begin{document}

\newcommand{\be}{\begin{equation}}
\newcommand{\ee}{\end{equation}}
\newcommand{\bea}{\begin{eqnarray}}
\newcommand{\eea}{\end{eqnarray}}
\newcommand{\beaa}{\begin{eqnarray*}}
\newcommand{\eeaa}{\end{eqnarray*}}

\title{Message and state cooperation in multiple access channels}

\author{Haim Permuter,  Shlomo (Shitz) Shamai, and Anelia Somekh-Baruch  \\
\thanks{H. Permuter is with the department of Electrical and Computer Engineering,
 Ben-Gurion University of the Negev, Beer-Sheva, Israel (haimp@bgu.ac.il).
S. (Shitz) Shamai is with the Department of Electrical Engineering,
Technion-Israel Institute of Technology, Haifa, Israel
(sshlomo@ee.technion.ac.il). A. Somekh-Baruch  is with the School of
Engineering, Bar-Ilan University, Ramat-Gan,  , Israel
(anelia.somekhbaruch@gmail.com).

This work has been supported by the CORNET Consortium sponsored by
the Chief Scientist of the Israel Ministry for Industry and
Commerce. }
}


%

\maketitle \vspace{-1.4cm}

\begin{abstract}
We investigate the capacity of a multiple access channel with
cooperating encoders where partial state information is known to
each encoder and full state information is known to the decoder. The
cooperation between the encoders has a two-fold purpose: to generate
empirical state coordination between the encoders, and to share
information about the private messages that each encoder has. For
two-way cooperation, this two-fold purpose is achieved by
double-binning, where the first layer of binning is used to generate
the state coordination similarly to the two-way source coding, and
the second layer of binning is used to transmit information about
the private messages. The complete result provides the framework and
perspective for addressing a complex level of cooperation that mixes
states and messages in an optimal way.
\end{abstract}

\vspace{-0.0cm}
\section{Introduction}
State-dependent channels describe a rich variety of communication
models spanning the cases, where the states are governed by physical
phenomena (such as fading), and accounting also for situations where
the states model effects of interfering transmissions.
\begin{figure}[h!]{
\psfrag{B}[][][1]{Encoder1} \psfrag{D}[][][1]{Encoder2}
\psfrag{m1}[][][1]{$m_1 \ \ \ \ $}
\psfrag{m2}[][][1]{$\in\{1,...,2^{nR_1}\}\ \ \ \ \ \ \ $}
\psfrag{m3}[][][1]{$m_2 \ \ \ \ $}
\psfrag{m4}[][][1]{$\in\{1,...,2^{nR_2}\}\ \ \ \ \ \ \ $}
\psfrag{P}[][][1]{$P(y|x_1,x_2,s_1,s_2)$} \psfrag{x1}[][][1]{$\ \
X_1$} \psfrag{x2}[][][1]{$\ \ \ X_2$} \psfrag{M}[][][1]{MAC}
\psfrag{s2}[][][1]{$S_2$} \psfrag{s1}[][][1]{$S_1$}
\psfrag{s12}[][][1]{$S_1,S_2\ \ \ \ $}\psfrag{Yi}[][][1]{$Y$}
\psfrag{W}[][][1]{Decoder} \psfrag{t}[][][1]{$S_1,S_2 \sim
P(s_1,s_2)$}

\psfrag{a}[][][1]{a} \psfrag{b}[][][1]{b}
\psfrag{c12}[][][1]{$C_{21}$} \psfrag{c12b}[][][1]{$C_{12} \ \ \ \
$}

\psfrag{Y}[][][1]{$\ \ \ \hat m_1$} \psfrag{R}[][][1]{$R_s\ \ \ \ \
\ \ $} \psfrag{Y2}[][][1]{$\  \hat m_2$}

\psfrag{A1}[][][1]{State} \psfrag{A2}[][][1]{Encoder}

\centerline{\includegraphics[width=14cm]{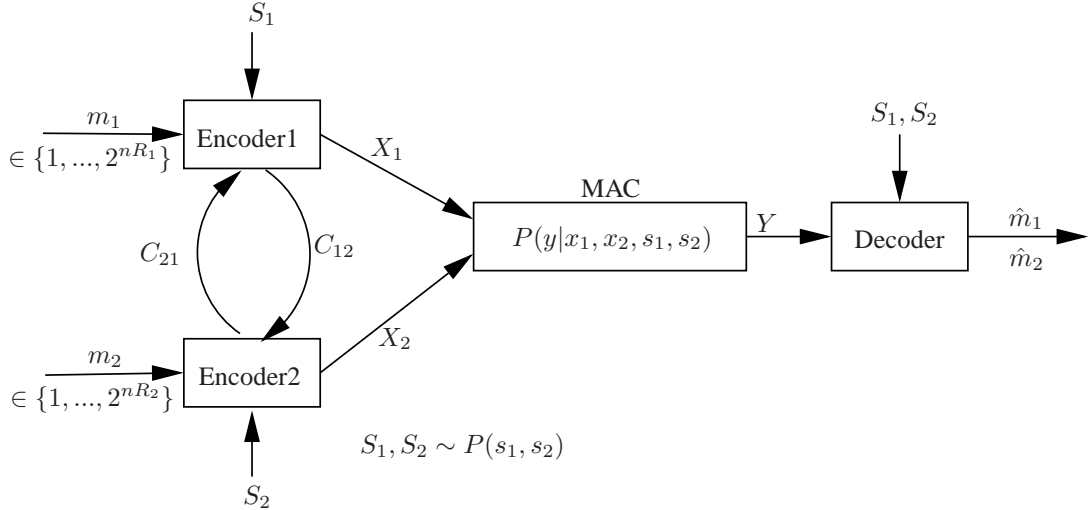}}
\caption{MAC with cooperation where different partial state
information is known to each encoder, and full state information is
known to the decoder.}\label{f_mac_coperating}
 }\end{figure} Their wide
applicability, theoretical importance, and practical implications,
led to intensive information theoretic studies. We focus here on a
multiple-access channel (MAC), where the channel is affected by the
state $(S_1,S_2)$ known partly at the transmitters. That is, state
$S_1$ is available at Transmitter 1, while $S_2$ is known at
Transmitter 2. This can be associated with local cognition, that is,
Transmitter 1 learns before hand about the sequence {$S_1$}, while
Transmitter 2 learns about {$S_2$}. We further assume that the
states, which can be viewed as channel-affecting parameters, are
known at the receiving point, or alternatively are retrieved
accurately by the receiver. This is a standard problem, which falls
within the class of decentralized processing at the transmitters.
The focus of this work is the implications of transmitter
cooperation facilitated by an orthogonal finite capacity link. This
link can be used both to share state information, as to facilitate a
more coordinated operation, up to a degree of central coordination,
achieved when both transmitters know accurately $(S_1,S_2)$. The
cooperation link can also be employed to share messages, to the
extreme of full message cooperation, turning the problem into a
single two-elements (antennas) transmitter. The interplay among
these types of cooperation is at the center of our paper, and here
the optimal approach, given in terms of the associated capacity
region, is found. Evidently the derivation of this general result is
extending
previous important cases as it is detailed in the following. 

Willems \cite{willems82_dissertation, Willems83_cooperating}
introduced and derived the capacity region of the multiple access
channel (MAC) with cooperating encoders. He  showed that to achieve
the capacity region the encoders should use the cooperation link in
order to share parts of their private messages and then use a coding
scheme for the ordinary MAC, which was found earlier by  Slepian and
Wolf
\cite{Slepian_Wolf_MAC73}.

In this paper, we consider the problem of MAC with cooperating
encoder, where different partial state information is known at each
encoder and perfect state information is known at the decoder. The
setting of the problem is depicted in Fig. \ref{f_mac_coperating}.
The state of the channel is given by the pair $(S_1,S_2)$, where
Encoder 1 knows $S_1$, Encoder 2 knows $S_2$, and the decoder knows
the pair $(S_1,S_2)$. The cooperation links $C_{12}$ and $C_{21}$
may increase the capacity region  by transmission of the state
information that is missing to the encoders and by sharing parts of
the private messages $(m_1,m_2)$. Here the transmission of the state
information is done by achieving an empirical coordination
\cite{CuffPermuterCover09SubmittedIT} of the state information,
namely, generating sequences of action  that are functions of the
cooperation and are jointly typical with the state information.
Simultaneously, these sequences of action are designed in such a way
that they  allow the encoders to share parts of their private
messages. To achieve this purpose we use double-binning, a technique
that was used by  Liu et. al \cite{LiuMaric_double_binning_08,
LiuPoor_09} for achieving secrecy capacity in the broadcast channel.

The problem of cooperating encoders with partial state information
combines two kinds of settings that are widely treated in the
literature; the first is limited-rate noise-free cooperation between
users and the second is limited-rate noise-free state information
that is available to encoders/decoders.

Cooperation between users through a noise-free limited-rate link has
been  investigated in various of multi-user settings such as in MAC
\cite{willems82_dissertation, Willems83_cooperating,
Gunduz08_cooperating_MAC_compound, Wigger_gaussian_cop},
interference channel
\cite{Viswanath09_destination_intereferncechannel_cooperation,Viswannath09_intereferncechannel_cooperation09,
WangTse_interference_cooperation09,
Bagheri_interference_cooperation09, NgJGM07_intereference_coope,
MaricKramer07_interegernce, Maric_inreference_cooperating08},
broadcast channel \cite{DaboraServetto_cooperative_BC08}, relay
channels
\cite{GunduzE07_cooperative_relaying,LalithaSankarKramer_Relay_Vs_Coope,GoldsmithNg_relay_coope_wireless},
and cellular networks
\cite{SimeoneSomekhShamai09_cooperative_cellular_network}. A
comprehensive survey of cooperation and its role in communication is
given in  \cite{KramerMaric_monoigraph_cooperation06}. Recently,
cooperation between encoders where state information is available
was  considered in
\cite{HaghiAref10_coopertaive_mac_state_fading_isit,HaghiAref10_coopertaive_mac_state_fading_IT}
where it is assumed that the cooperation is allowed only before the
state information is available at the encoders. In this paper, we
take a different approach, assuming that the cooperation occurs
after the state information becomes available, the cooperation may
include parts of the private message and the state information as
well.

The second setting, that is, limited-rate state information at
encoders/decoders,  was first treated by Heegard and El-Gamal
\cite{HeegardElGamal_state_encoded83}. The case, most related to the
setting in this paper, where full state information is available at
the decoder and limited-rate state information is known at the
encoder was solved by Cemal and Steinberg for the point-to-point
channel \cite{SteinbergCemal07_state_encoded} and for the MAC
\cite{Steinberg_Cemal_MAC05}. The main difference between the
setting here and the setting in \cite{Steinberg_Cemal_MAC05} is that
here the limited-rate encoder knows the state and the private
message rather than just the private message; therefore, a scheme
which combines message information and state information is needed.

The remainder of the paper is organized as follows. In Section
\ref{s_one_way_cop}, we derive the capacity region where only one
cooperation link from Encoder 1 to Encoder 2 exists. This setting
helps us to gain the intuition necessary for solving the extended
problem of two-way cooperation, which is solved in Section
\ref{s_two_way_cop}. In Section \ref{s_example}, we solve a specific
example and compare the capacity region to two different cooperation
settings given in
\cite{HaghiAref10_coopertaive_mac_state_fading_isit} and in
\cite{Steinberg_Cemal_MAC05}. In addition, in Section
\ref{s_example}, we check the strategy of splitting the cooperation
link into message-only link and state-only link, and we show that
this naive strategy is strictly suboptimal.

\section{One-way cooperation\label{s_one_way_cop}}
In this section, we consider a special case, in which there is only
one-way cooperation from Encoder 1 to Encoder 2. In addition,  we
assume that Encoder 1 and the decoder have full non-causal state
information.
\begin{figure}[h!]{
\psfrag{B}[][][1]{Encoder1} \psfrag{D}[][][1]{Encoder2}
\psfrag{m1}[][][1]{$m_1 \ \ \ \ $}
\psfrag{m2}[][][1]{$\in\{1,...,2^{nR_1}\}\ \ \ \ \ \ \ $}
\psfrag{m3}[][][1]{$m_2 \ \ \ \ $}
\psfrag{m4}[][][1]{$\in\{1,...,2^{nR_2}\}\ \ \ \ \ \ \ $}
\psfrag{P}[][][1]{$P_{Y|X_1,X_2,S}$} \psfrag{x1}[][][1]{$\ \ \ \ \ \
\ \ \ \ \ \ X_1(m_1,S^n)$} \psfrag{x2}[][][1]{$\; \ \ \ \ \ \ \ \ \
\ \ \ \  X_2(m_2,m_{12})$} \psfrag{M}[][][1]{MAC}
\psfrag{s}[][][1]{$S$} \psfrag{Yi}[][][1]{$Y$}
\psfrag{W}[][][1]{Decoder} \psfrag{t}[][][1]{$$}

\psfrag{a}[][][1]{a} \psfrag{b}[][][1]{b} \psfrag{c12}[][][1]{$
m_{12}\in$} \psfrag{c12b}[][][1]{$ \ \ \ \ \ \
\{1,...,2^{nC_{12}}\}$}

\psfrag{Y}[][][1]{$\ \ \ \ \ \ \ \ \ \ \ \hat m_1(Y^n,S^n)$}
\psfrag{Y2}[][][1]{$\ \ \ \ \ \ \ \ \ \ \hat m_2(Y^n,S^n)$}

\centerline{\includegraphics[width=14cm]{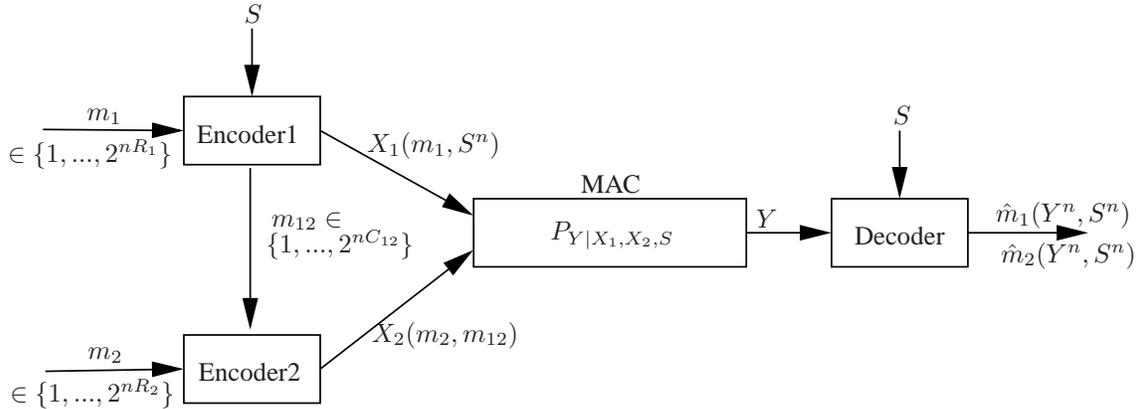}}
\caption{MAC with one-way conferencing and state information at one
encoder and the decoder}\label{f_mac_one_way_cop}
 }\end{figure}
This setting captures the idea of, simultaneously, sharing a part of
the  private message $m_1$ and sharing the information on channel
state $S$. The setting is depicted in Fig. \ref{f_mac_one_way_cop}.
We start by defining the notation and the code for this setting,
then we state the capacity region, explain the intuition and provide
its proof.

The MAC setting consists of two transmitters (encoders) and one
receiver (decoder). Each sender $l\in\{1,2\}$ chooses an index $m_l$
uniformly from the set $\{1,...,2^{nR_l}\}$ and independently of the
other sender. The input to the channel from encoder $l\in\{1,2\}$ is
denoted by $\{X_{l,1},X_{l,2},X_{l,3},...\}$, and the output of the
channel is denoted by $\{Y_1,Y_2,Y_3,...\}$. The state at time $i$,
i.e., $S_i\in \mathcal S$, takes values in a finite set of possible
states $\mathcal S$. The channel is characterized by a conditional
probability $P(y_i|x_{1,i},x_{2,i},s_{i})$ and by the state
probability $P(s_i)$. Both probabilities do not depend on the time
index $i$ and satisfy
\begin{equation}
P(y_i,s_{i+1}|x_1^i,x_2^i,s^{i},y^{i-1})=P(y_i|x_{1,i},x_{2,i},s_{i})P(s_i),
\end{equation}
where the superscripts denote sequences in the following way:
$x_l^i=(x_{l,1},x_{l,2},...,x_{l,i}), \; l\in\{1,2\}$.

\begin{definition}\label{def_one_way}
A $(2^{nR_1},2^{nR_2},2^{nC_{12}},n)$ {\it code} with one-way
cooperating encoder as shown in Fig. \ref{f_mac_one_way_cop}
consists of three encoding functions
\begin{eqnarray}
f_1&:&\{1,...,2^{nR_1}\}\times \mathcal S^n \mapsto \mathcal
X_1^n,\nonumber \\
f_{12}&:&\{1,...,2^{nR_1}\}\times \mathcal S^n \mapsto \mathcal
\{1,...,2^{nC_{12}}\},\nonumber \\
f_2&:&\{1,...,2^{nR_2}\}\times \{1,...,2^{nC_{12}}\} \mapsto
\mathcal X_2^n,
\end{eqnarray}
and a decoding function,
\begin{equation}
g:\mathcal Y^n \times \mathcal S^n \mapsto \{1,...,2^{nR_1}\} \times
\{1,...,2^{nR_2}\}.
\end{equation}
\end{definition}

The {\it average probability of error} for
$(2^{nR_1},2^{nR_2},2^{nC_{12}},n)$ code is defined as
\begin{equation}
P_e^{(n)}=\frac{1}{2^{n(R_1+R_2)}} \sum_{m_1,m_2}
\Pr\{g(Y^n,S^n)\neq(m_1,m_2)|(m_1,m_2) \text{ sent}\}.
\end{equation}
A rate $(R_1,R_2)$ is said to be {\it achievable} for the one-way
cooperating MAC with cooperation link $C_{12},$ if there exists a
sequence of $(2^{nR_1},2^{nR_2},2^{nC_{12}},n)$ codes with
$P_e^{(n)}\to 0$. The {\it capacity region} of MAC is the closure of
all achievable rates. The following theorem describes the capacity
region of one-way cooperating MAC.
\begin{theorem}\label{t_mac_one_way_cop}
The capacity region of the MAC with a cooperating encoder that has state
information  as shown in Fig. \ref{f_mac_one_way_cop} is the closure of the set that contains all rates that satisfy
\begin{eqnarray}
C_{12}&\geq& I(U;S) \label{e_c1} \\
R_1&\leq & I(X_1;Y|X_2,S,U)+C_{12}-I(U;S)\\
R_2&\leq & I(X_2;Y|X_1,S,U)  \\
R_1+R_2&\leq& \min \left\{ \begin{array}{c}
I(X_1,X_2;Y|S,U)+C_{12}-I(U;S),\\ I(X_1,X_2;Y|S) \end{array}
\right\},\label{e_R1+R2}
\end{eqnarray}
for some joint distribution of the form
\begin{equation}\label{e_c5}
P(s)P(u,x_1|s)P(x_2|u)P(y|x_1,x_2,s).
\end{equation}

\end{theorem}

\begin{lemma}
\label{l_properties_R_one_way_cop}
\begin{enumerate}
\item \label{lemma_properties_R_convex_one_way_cop}
The capacity region described in Theorem \ref{t_mac_one_way_cop}, given in (\ref{e_c1})-(\ref{e_c5}), is convex.
\item \label{lemma_properties_R_size_one_way_cop}
It is enough to restrict the
alphabet of the auxiliary random variable $U$ in Theorem \ref{t_mac_one_way_cop}  to satisfy
\begin{eqnarray}
|\mathcal U|\leq\min(|\mathcal X_1||\mathcal X_2||\mathcal S|+3, |\mathcal Y||\mathcal S|+4).
\end{eqnarray}
\end{enumerate}
\end{lemma}

Before proving the theorem and the lemma let us investigate the role
of the auxiliary random variable $U$ in Theorem
\ref{t_mac_one_way_cop}. The random variable $U$  plays a double
role: first, it generates an empirical coordination between the two
encoders regarding the state of the channel; second, it generates a
common message between the two encoders. Let us look at two special
cases which emphasize these two roles.

{\it Case 1: The point-to-point case \cite{SteinbergCemal07_state_encoded}, i.e., $R_1=0$ and $P(y|x_1,x_2,s)=P(y|x_2,s)$.} For this case the rate region of Theorem \ref{t_mac_one_way_cop} becomes
\begin{eqnarray}
C_{12}&\geq& I(U;S) \\
R_2&\leq & I(X_2;Y|S,U)  \\
R_2&\leq& \min \left\{ \begin{array}{c}
I(X_2;Y|S,U)+C_{12}-I(U;S)\\ I(X_2;Y|S) \end{array}
\right\},
\end{eqnarray}
 which is simply
\begin{eqnarray}
C_{12}&\geq& I(U;S)  \\
R_2&\leq & I(X_2;Y|S,U)  \\
\end{eqnarray}
for a joint distribution of the form $P(s)P(u|s)P(x_2|u)P(y|x_2,s)$.

{\it Case 2: $ |\mathcal S|=1$, the memoryless case \cite{Willems83_cooperating}.}
 In this case $I(U;S)=0$, hence we obtain a special case of MAC with cooperation and the rate region of Theorem \ref{t_mac_one_way_cop} becomes
\begin{eqnarray}
R_1&\leq & I(X_1;Y|X_2,U)+C_{12}\\
R_2&\leq & I(X_2;Y|X_1,U)  \\
R_1+R_2&\leq& \min \left\{ \begin{array}{c} I(X_1,X_2;Y|U)+C_{12}\\
I(X_1,X_2;Y) \end{array} \right\},
\end{eqnarray}
for a joint distribution of the form
$P(u)P(x_1|u)P(x_2|u)P(y|x_2,x_1)$.

\begin{figure}[h!]{
\psfrag{box1}[][][1]{Encoder}
\psfrag{box1}[][][1]{Encoder}
\psfrag{b1}[][][1]{$S^n, M\in\{1,...,2^{nR_m}\}\ \ \ \ \ \ \ \ \ \ \ \ $}
\psfrag{t}[][][1]{$S^n$ is  i.i.d., $S\sim P(s)$\ \ \ \ }
\psfrag{a2}[][][1]{$T(S^n,m)$}
\psfrag{t1}[][][1]{$\in\{1,...,2^{nR}\}$}
\psfrag{b3}[][][1]{$U^n(T)$}
\psfrag{t2}[][][1]{$\hat M=g(U^n)$, $\lim_{n\to \infty} \Pr\{g(U^n)\neq  M\}=0$}
\psfrag{t3}[][][1]{$\lim_{n\to \infty} \Pr\{(U^n,S^n)\in T_{\epsilon}^{(n)}(U,S)\}=1$}

\psfrag{box2}[][][1]{decoder}
\centerline{\includegraphics[width=12cm]{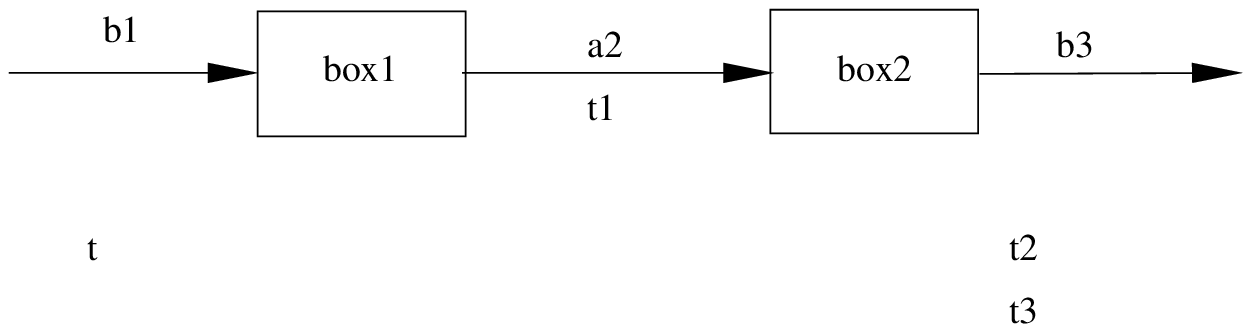}}
\caption{A problem that illustrates the double role of cooperation.
One one hand, the sequence $U^n$ needs to be jointly typical with
$S^n$, and on the other hand, one should be able to reconstruct the
message $m$ with high probability. 
}\label{f_double_role}
 }\end{figure}

Note that in the first case the role  of the auxiliary random
variable $U$ is to generate an empirical coordination $P_{U|S}$, and
then use the sequence $U^n$ as common side information at the
encoder and decoder. In the second case, the auxiliary random
variable represents the common message $m_0$ between the two
encoders, and the decoder needs to decode it. In Theorem
\ref{t_mac_one_way_cop}, these two roles are combined. Namely, the
sequence $U^n$ needs to be coordinated with $S^n$ and simultaneously
represents a common message.  Fig. \ref{f_double_role} illustrates
the role of cooperation. On one hand, the cooperation needs to
generate a sequence $U^n$ that is jointly typical with $S^n$, i.e.,
$\lim_{n\to \infty} \Pr\{(U^n,S^n)\in T_{\epsilon}^{(n)}(U,S)\}=1$,
and on the other hand, there should be a function $g(U^n)$ such that
one can estimate the message $M$ with high probability, i.e.,
$\lim_{n\to \infty} \Pr\{g(U^n)\neq  M\}=0$. If $R>R_m+I(U;S)$ and
$H(U|S)\geq R_m$, this goal can be achieved.
Combining these two roles (generating empirical coordination and
transmitting a message) is done by binning, where the bin number
represents the common message and in each bin there will be enough
codewords $U^n$ such that at least one codeword is jointly typical
with $S^n$. This is similar to the role of the auxiliary random
variable in Gelfand-Pinsker \cite{GePi80}, where the sequence of the
auxiliary random variables that is generated needs to represent a
message that is transmitted via the channel and needs to be jointly
typical with the sequence of the channel states.

Next we present a formal proof of Theorem 1. Throughout the
achievability proofs in the paper we use the definition of a strong
typical set.  The set $T^{(n)}_\epsilon(X,Y,Z)$ of
$\epsilon$-typical $n-$sequences is defined by
$\{(x^n,y^n,z^n):\frac{1}{n}N(x,y,z|x^n,y^n,z^n)-p(x,y,z)|\leq
\epsilon p(x,y,z) \forall (x,y,z)\in \mathcal X\times \mathcal
Y\times \mathcal Z\}$, where $N(x,y,z|x^n,y^n,z^n)$ is the number of
appearances of $(x,y,z)$ in the $n-$sequnce $(x^n,y^n,z^n)$.
Furthermore, we will use the following well-known lemma
\cite{Csiszar81,CovThom06,Kramer07_book_multi_user_IT,ElGamalKim_lecturenotes10},
\begin{lemma}[Joint typicality lemma]\label{l_typical}
Consider a joint distribution $P_{X,Y,Z}$ and suppose $(x^n,y^n)\in
T_\epsilon^{(n)}(X,Y)$. Let $\tilde Z^n$ be distributed according to
$\prod_{i=1}^n P_{Z|X}(\tilde z_i|x_i)$. Then,
\begin{equation}
\Pr\{(x^n,y^n,\tilde Z^n)\in T^{(n)}_{\epsilon}(X,Y,Z)\}\leq
2^{-n(I(Y;Z|X)-\delta(\epsilon))},
 \end{equation}
 where $\lim_{\epsilon\to 0}\delta(\epsilon)= 0$.
\end{lemma}

{\it Proof of Theorem \ref{t_mac_one_way_cop}:}  {\bf Achievability
part.}

 {\it Code construction:} Generate $2^{nC_{12}}$ codewords
$U^n$ independently using i.i.d. $\sim P(u)$, and assign them into
$2^{n(C_{12}-I(U;S)-\epsilon)}$ bins. Hence, in each bin there are
$2^{n(I(U;S)+\epsilon)}$ codewords. For each codeword $u^n(j)$,
where $j=1,2,...,2^{nC_{12}}$  and for each $s^n\in \mathcal S^n$
generate $2^{n(R_1-(C_{12}-I(U;S)-\epsilon))}$ codewords $X_1^n$
according to i.i.d. $\sim P(x_1|u,s)$ and for each $u^n(j)$, where
$j=1,2,...,2^{nC_{12}}$, generate $2^{n(R_2-(C_{12}-I(U;S)))}$
codewords $X_2^n$ according to i.i.d. $\sim P(x_2|u)$.

{\it Encoder:} Split  message $m_1\in [1,...,2^{nR_1}]$ into two
messages $m_{1,a}\in [1,...,2^{n(C_{12}-I(U;S)-\epsilon)}]$ and
$m_{1,b}\in [1,...,2^{n(R_1-(C_{12}-I(U;S)-\epsilon))}]$. 
Now, associate each message $m_{1,a}\in[1,...,
2^{n(C_{12}-I(U;S)-\epsilon)}]$ with a bin, where in each bin there
are $2^{n(I(U;S)+\epsilon)}$  codewords $u^n$, indexed by $l\in
[1,...,2^{n(I(U;S)+\epsilon)}]$. Find in the chosen bin a codeword,
denoted by $u^n(m_{1,a},s^n)$, with the smallest lexicographical
order that is jointly typical with $s^n$  and send its index
$[1,...,2^{C_{12}}]$ to Encoder 2. If such a codeword $u^n$ does not
exist, namely, among the codewords in the bin none is jointly
typical with $s^n$, choose an arbitrary $u^n$ from the bin (in such
a case the decoder will declare an error). Now, Encoder 1 transmits
$x_1^n(s^n,u^n(m_{1,a},s^n),m_{1,b})$, and Encoder 2 transmits
$x_2^n(u^n(m_{1,a},s^n),m_{2})$.

{\it Decoder:} The decoder knows $s^n$ and $y^n$ and looks for the
indices $\hat m_{1,a}\in [1,...,2^{n(C_{12}-I(U;S)-\epsilon)}]$,
$\hat m_{1,b}\in[1,..., 2^{n(R_1-(C_{12}-I(U;S)-\epsilon))}]$, $\hat
m_2\in [1,...,2^{nR_2}]$ such that
\begin{equation}
\left(u^n(\hat m_{1,a},s^n),x_1^n(s^n,u^n(\hat m_{1,a},s^n),\hat
m_{1,b}),x_2^n(u^n(\hat m_{1,a},s^n),\hat m_{2}),s^n,y^n\right)\in
T_{\epsilon}^{(n)}(U,X_1,X_2,S,Y),
\end{equation}
If none or more than one
such triplet is found, an error is declared. The estimated message sent
from Encoder 1 is $(\hat m_{1,a},\hat m_{1,b})$, and the estimated
message transmitted from Encoder 2 is $\hat m_2$.

{\it Error analysis:} 
Assume $(m_{1,a},m_{1,b},m_2)=(1,1,1)$. Let us define the event
\begin{equation}
E_{i,j,k}\triangleq\left\{
\left(u^n(i,s^n),x_1^n(s^n,u^n(i,s^n),j),x_2^n(u^n(i,s^n),j),s^n,y^n\right)\in
T_{\epsilon}^{(n)}(U,X_1,X_2,S,Y)\right\}.
\end{equation}
An error occurs if either the correct codewords are not jointly
typical with the received sequences, i.e., $E_{1,1,1}^c$, or there
exists a different $(i, j,k)\neq(1,1,1)$ such that $E_{i,j,k}$
occurs. From the union of bounds we obtain that
\begin{equation}\label{e_pe}
P_e^{(n)}\leq \Pr(E_{1,1,1}^c)+\sum_{i=1,j=1,k>1}
\Pr(E_{i,j,k})+\sum_{i=1,j>1,k=1} \Pr(E_{i,j,k})+\sum_{i=1,j>1,k>1}
\Pr(E_{i,j,k})+\sum_{i>1,j\geq1,k\geq 1} \Pr(E_{i,j,k}).
\end{equation}
Now let us show that each term in (\ref{e_pe}) goes to zero as the
blocklength of the code $n$ goes to infinity.
\begin{itemize}
\item Upper-bounding $\Pr(E_{1,1,1}^c)$:
Since the number of codewords in each bin is larger than
$2^{nI(U;S)}$, and since the codewords were generated i.i.d., with
high probability there will be at least one codeword that is jointly
typical with $s^n$. We denote this sequence as $u^n(1)$.
Furthermore, given that $(u^n(1),s^n)\in T^{(n)}_\epsilon (U,S)$, it
follows from the law of large numbers that $\Pr(E_{1,1,1}^c)\to 0$
as $n$ goes to infinity.
\item Upper-bounding $\sum_{i=1,j=1,k>1}
\Pr(E_{i,j,k})$:  The probability that $Y^n$, which is generated
according to $P(y|x_1,s,u)$, is jointly typical with $x_2^n$, which
was generated according to $P(x_2|u)=P(x_2|u,s,x_1)$, where
$(x_1^n,s^n,u^n)\in T^{(n)}_\epsilon(X_1,S,U)$ is bounded by (Lemma
\ref{l_typical})
\begin{equation}
\Pr\{(x_1^n,X_2^n,u^n,s^n,Y^n)\in
T^{(n)}_\epsilon|(x_1^n,u^n,s^n)\in T^{(n)}_\epsilon\}\leq
2^{-n(I(X_2;Y|X_1,S,U)-\delta(\epsilon))}.
\end{equation}
Hence, we obtain
\begin{eqnarray}\label{e_u1}
\sum_{i=1,j=1,k>1}
\Pr(E_{i,j,k})&\leq&2^{nR_2}2^{-n(I(X_2;Y|X_1,S,U)-\delta(\epsilon))}
\end{eqnarray}

\item Upper-bounding $\sum_{i=1,j>1,k=1} \Pr(E_{i,j,k})$:
The probability that $Y^n$ which is generated according to
$P(y|x_2,s,u)$ is jointly typical with $x_1^n$ which was generated
according to $P(x_1|u,s)=P(x_1|u,s,x_2)$, where $(x_2^n,s^n,u^n)\in
T^{(n)}_\epsilon(X_2,S,U)$ is upper bounded by
$2^{-n(I(X_1;Y|X_2,S,U)-\delta(\epsilon))}$, hence
\begin{eqnarray}
\sum_{i=1,j>1,k=1}
\Pr(E_{i,j,k})&\leq&2^{n(R_1-(C_{12}-I(U;S)-\epsilon))}
2^{-n(I(X_2;Y|X_1,S,U)-\delta(\epsilon))}.
\end{eqnarray}

\item  Upper-bounding $\sum_{i=1,j>1,k>1}
\Pr(E_{i,j,k})$
\begin{eqnarray}
\sum_{i=1,j>1,k>1}
\Pr(E_{i,j,k})&\leq&2^{n(R_2+R_1-(C_{12}-I(U;S)-\epsilon))}
2^{-n(I(X_2,X_1;Y|S,U)-\delta(\epsilon))}.
\end{eqnarray}

\item  Upper-bounding $\sum_{i>1,j\geq1,k\geq 1} \Pr(E_{i,j,k})$

\begin{eqnarray}\label{e_u5}
\sum_{i>1,j\geq1,k\geq 1}
\Pr(E_{i,j,k})&\leq&2^{n(C_{12}-I(U;S)-\epsilon)}2^{n(R_1-(C_{12}-I(U;S)-\epsilon))}2^{nR_2}
2^{-n(I(X_2,X_1,U;Y|S)-\delta(\epsilon))}\nonumber\\
&=&2^{n(R_1+R_2-I(X_2,X_1,U;Y|S)-\delta(\epsilon))}
\end{eqnarray}

\end{itemize}

Therefore, combining the upper bounds (\ref{e_u1})-(\ref{e_u5}) into
(\ref{e_pe}), we obtain that if rate-pair $(R_1,R_2)$ is inside the
rate region given by (\ref{e_c1})-(\ref{e_c5}), then there exists a
sequence of codes $(2^{nR_1},2^{nR_2},2^{nC_{12}},n)$ such that
$P_{\epsilon}^{(n)}$ goes to zero as $n\to \infty$.


{\bf Converse part:} Assume that we have a
$(2^{nR_1},2^{nR_2},2^{nC_{12}},n)$ code as in Definition
\ref{def_one_way}. We will show the existence of a joint
distribution $P(s)P(u|s)P(x_1|s,u)P(x_2|u)P(y|x_1,x_2)$ that
satisfies~(\ref{e_c1})-(\ref{e_R1+R2}) within some $\epsilon_n$,
where  $\epsilon_n$ goes to zero as $n\to\infty$. Denote
$M_{12}=f_{12}(M_1,S^n)$. Then,
\begin{eqnarray}
nC_{12}&\geq& H(M_{12})\nonumber \\
&\geq& I(M_{12};S^n)\nonumber \\
&\stackrel{(a)}{=}& \sum_{i=1}^n I(S_i;M_{12},S^{i-1})\nonumber \\
&\stackrel{(b)}{=}& \sum_{i=1}^n I(S_i;U_i),\label{e_nc12_one}
\end{eqnarray}
where (a) follows from the fact that $S_i$ is i.i.d. and (b) follows from the definition of $U_i$, which is
\begin{equation}\label{e_def_Ui}
U_i \triangleq (M_{12},S^{i-1}).
\end{equation}
Next, consider
\begin{eqnarray}\label{e_Hm1all}
nR_1&=& H(M_{1})\nonumber \\
&=& H(M_{1}|S^n,M_2)\nonumber \\
&=& H(M_{1},M_{12}|S^n,M_2)\nonumber \\
&=& H(M_{12}|S^n,M_2)+H(M_{1}|S^n,M_2,M_{12})\nonumber \\
&\leq & H(M_{12}|S^n)+H(M_{1}|S^n,M_2,M_{12}).
\end{eqnarray}
Now, let us consider the terms $H(M_{12}|S^n)$ and
$H(M_{1}|S^n,M_2,M_{1,2})$ separately.
\begin{eqnarray}\label{e_HM12}
H(M_{12}|S^n)
&=& H(M_{12}|S^n)-H(M_{12})+H(M_{12})\nonumber \\
&\leq & nC_{12}-I(S^n;M_{12})\nonumber \\
&=& nC_{12}-\sum_{i=1}^n I(S_i;U_i),
\end{eqnarray}
where the last equality follows from  (\ref{e_nc12_one}) where it is
shown that $I(M_{12};S^n)=\sum_{i=1}^n  I(S_i;U_i)$. Further,
\begin{eqnarray}\label{e_Hm1}
\lefteqn{H(M_{1}|S^n,M_2,M_{12})}\nonumber \\
&\stackrel{(a)}{=}& I(M_1;Y^n|S^n,M_2,M_{12})+n\epsilon_n \nonumber \\
&=& H(Y^n|S^n,M_2,M_{12})-H(Y^n|S^n,M_2,M_{12},M_1)+n\epsilon_n \nonumber \\
&\stackrel{(b)}{=}& H(Y^n|S^n,X_2^n,M_2,M_{12})-H(Y^n|S^n,X_2^n,X_1^n,M_2,M_{12},M_1)+n\epsilon_n \nonumber \\
&=& \sum_{i=1}^n H(Y_i|Y^{i-1},S^n,X_2^n,M_2,M_{12})-H(Y_i|S^n,X_2^n,X_1^n,M_2,M_{12},M_1,Y^{i-1})+n\epsilon_n \nonumber \\
&\stackrel{(c)}{\leq}& \sum_{i=1}^n
H(Y_i|S_i,X_{2,i},M_{12},S^{i-1})-H(Y_i|S_i,X_{2,i},X_{1,i},M_{12},S^{i-1})+n\epsilon_n,
\end{eqnarray}
where (a) follows from Fano's inequality and from the definition
$\epsilon_n\triangleq R_1P_e^{(n)}$, (b) follows from the fact that
$X_1^n$ is a deterministic function of $(S^n,M_1)$ and $X_2^n$ is a
deterministic function of $(M_2,M_{12})$,  and (c) from the fact
that conditioning reduces entropy and from the Markov chain
$Y_i-(S_i,X_{2,i},X_{1,i})-(S^n,X_2^n,X_1^n,M_2,M_{12},M_1)$.
Substituting Inequalities (\ref{e_HM12}) and (\ref{e_Hm1}) into
(\ref{e_Hm1all}), we obtain
\begin{eqnarray}
nR_1&\leq& \sum_{i=1}^n
I(Y_i;X_{1,i}|S_i,X_{2,i},U_i)-I(S_i;U_i)+nC_{12}.
\end{eqnarray}
Similarly, we have
\begin{eqnarray}\label{e_Hm1all}
nR_2&=& H(M_{2})\nonumber \\
&=& H(M_{2}|S^n,M_1,M_{12})\nonumber \\
&\leq& \sum_{i=1}^n I(Y_i;X_{2,i}|S_i,X_{1,i},U_i),
\end{eqnarray}
where the last inequality follows from similar steps as in
(\ref{e_Hm1}). Regarding the sum-rate we have
\begin{eqnarray}
nR_1+nR_2&=&H(M_1,M_2)\nonumber \\
&=&H(M_1,M_2|S^n)\nonumber \\
&=&I(M_1,M_2;Y^n|S^n)+n\epsilon_n \nonumber \\
&=&\sum_{i=1}^n H(Y_i|Y^{i-1},S^n)-H(Y_i|M_1,M_2,S^n,X_1^n,X_2^n)+n\epsilon_n \nonumber \\
&\leq&\sum_{i=1}^n H(Y_i|S_i)-H(Y_i|S_i,X_{1,i},X_{2,i})+n\epsilon_n \nonumber \\
&\leq&\sum_{i=1}^n I(X_{1,i},X_{2,i};Y_i|S_i)+n\epsilon_n
\end{eqnarray}
and

\begin{eqnarray}
nR_1+nR_2&=&H(M_1,M_2)\nonumber \\
&=&H(M_1,M_2,M_{12}|S^n)\nonumber \\
&=&H(M_{12}|S^n)+H(M_1,M_2|M_{12},S^n)+n\epsilon_n,
\end{eqnarray}
and now using (\ref{e_HM12}) and similar steps as in (\ref{e_Hm1}) we obtain
\begin{eqnarray}
nR_1+nR_2&\leq&\sum_{i=1}^n   I(Y_i;X_{1,i},X_{2,i}|S_i,U_i)-I(S_i;U_i)+nC_{12}.
\end{eqnarray}

Now we verify that the  Markov chain $X_{2,i}-U_i-(X_{1,i},S_i)$
holds (this is due to  the Markov chain
$M_2-(M_{12},S^{i-1})-(M_1,S^n)$). Finally, let $Q$ be a random
variable independent of $(X^n,S^n,Y^n)$, and uniformly distributed
over the set $\{1,2,3,..,n\}$. Define the random variables
$U\triangleq(Q,U_Q)$. Using the simple observation that
$I(X_1,X_2;Y|S,Q)\leq I(X_1,X_2;Y|S)$, we obtain that the region
given in (\ref{e_c1})-(\ref{e_c5}) is an outer bound to any
achievable rate.
 \hfill\QED

{\it Proof of Lemma \ref{l_properties_R_one_way_cop}: } First we
prove that the capacity region described in Theorem
\ref{t_mac_one_way_cop}, (\ref{e_c1})-(\ref{e_c5}), is convex and
therefore there is no need to convexify it. Let $P_i$, $i=1,2,3$ be
three distributions of the form \begin{equation}\label{e_p_form}
P(s)P(u,x_1|s)P(x_2|u)P(y|x_1,x_2,s),
 \end{equation}
 which induce the quantities
 \begin{equation}\label{e_terms}
 (I_i(U;S), I_i(X_1;Y|X_2,S,U), I_i(X_2;Y|X_1,S,U), I_i(X_1,X_2;Y|S,U), I_i(X_1,X_2;Y|S)),
  \end{equation}
  for  $i=1,2,3$, respectively. In addition,  let $P_3=\alpha P_1+\overline \alpha P_2$, where $0\leq \alpha\leq 1$ and $\overline \alpha=1-\alpha$, furthermore when $q=1$ the distribution of $U,X_1,X_2$ is according to $P_1$ and when $q=2$ it is according to $P_2$. Let $Q$ be a binary random variable with $P(q=1)=\alpha$ and $P(q=2)=1-\alpha$. Let us denote $\tilde U=(U,Q)$, and note that $P_3$ is of the form of (\ref{e_p_form}) where $\tilde U$ replaces $U$. Finally, the convexity of the region in (\ref{e_c1})-(\ref{e_c5}) follows from the equalities $\alpha I_1(U;S)+\overline \alpha I_2(U;S)= I_3(\tilde U;S)$, and similar equalities for the other terms in (\ref{e_terms}), and from the inequality
  \begin{eqnarray}
  \alpha I_1(X_1,X_2;Y|S)+\overline \alpha  I_1(X_1,X_2;Y|S)&=&I_3(X_1,X_2;Y|S,Q)\nonumber \\
  &\leq&
  I_3(X_1,X_2;Y|S).
  \end{eqnarray}

Now, to prove the cardinality bound on $U$,  we invoke the support
lemma~\cite[p. 310]{Csiszar81}. The auxiliary random variable $U$
needs to have $|{\mathcal X_1}||{\mathcal X_2}||{\mathcal S}|-1$
letters to preserve $p(x_1,x_2,s)$
 plus four more to preserve the expressions
$H(S|U)$, $I(X_1;Y|X_2,S,U)$, $I_i(X_2;Y|X_1,S,U)$, and
$I(X_1,X_2;Y|S,U)$. Note that the joint distribution
$p(x_1,x_2,s,y)$ is preserved because of the Markov form
$U-(X_1,X_2,S)-Y$. Alternatively, the external random variable $U$
needs to have $|{\cal Y}||{\cal S}|-1$ letters to preserve $P(y,s)$
 plus five more to preserve the expressions
$H(S|U)$, $I(X_1;Y|X_2,S,U)$, $I_i(X_2;Y|X_1,S,U)$,  $I(X_1,X_2;Y|S,U)$, and $H(Y|X_1,X_2,S,U)$.
\hfill \QED

\section{Two-way cooperation\label{s_two_way_cop}}
Here we extend the setting from the previous section to a MAC with
 two-way cooperation  where  different state
information is available at each encoder and full state information
is available  at the receiver, as depicted in Fig.
\ref{f_mac_coperating}.

\begin{definition} \label{d_two_way} A $(2^{nR_1},2^{nR_2},2^{nC_{12}},2^{nC_{21}},n)$ {\it code} with two-way cooperating encoders, where each encoder has
partial state information,  consists of four encoding functions
\begin{eqnarray}
f_{12}&:&\{1,...,2^{nR_1}\}\times \mathcal S_1^n \mapsto \mathcal
\{1,...,2^{nC_{12}}\},\nonumber \\
f_{21}&:&\{1,...,2^{nR_2}\}\times \{1,...,2^{nC_{12}}\} \times
\mathcal S_2^n \mapsto
\{1,...,2^{nC_{21}}\},\nonumber \\
f_1&:&\{1,...,2^{nR_1}\}\times \{1,...,2^{nC_{21}}\}\times \mathcal
S_1^n \mapsto \mathcal
X_1^n,\nonumber \\
f_2&:&\{1,...,2^{nR_1}\}\times \{1,...,2^{nC_{12}}\}\times \mathcal
S_2^n  \mapsto \mathcal X_2^n,
\end{eqnarray}
and a decoding function,
\begin{equation}
g:\mathcal Y^n \times \mathcal S_1^n \times \mathcal S_2^n\mapsto
\{1,...,2^{nR_1}\} \times \{1,...,2^{nR_2}\}.
\end{equation}
\end{definition}
The probability of error, achievable rates, and the capacity region are defined similarly to  Definition \ref{def_one_way}. The next theorem states the capacity region of the two-way cooperating encoders with partial state information.

\begin{theorem}\label{t_two_way_cop}
The capacity region of the MAC with two-way cooperating encoders and
with partial state information  as shown in Fig.
\ref{f_mac_coperating} is the closure of the  set of rates that
satisfy
\begin{eqnarray}
C_{12}&\geq& I(U;S_1|S_2) \label{e_c12_lim_rate} \\
C_{21}&\geq& I(V;S_2|S_1,U) \label{e_c21_lim_rate} \\
R_1&\leq & I(X_1;Y|X_2,S_1,S_2,U,V)+C_{12}-I(U;S_1|S_2)\\
R_2&\leq & I(X_2;Y|X_1,S_1,S_2,U,V)+C_{21}-I(V;S_2|S_1,U)\\  \\
R_1+R_2&\leq& \min \left\{ \begin{array}{c}
I(X_1,X_2;Y|X_1,S_1,S_2,U,V)+C_{12}+C_{21}-I(U;S_1|S_2)-I(V;S_2|S_1,U)\\ I(X_1,X_2;Y|S_1,S_2) \end{array}
\right\},\label{e_R1+R2_lim_rate}
\end{eqnarray}
for some joint distribution of the form
\begin{equation}
P(s_1,s_2)P(u|s_1)P(v|s_2,u)P(x_1|s_1,u,v)P(x_2|s_2,u,v)P(y|x_1,x_2,s_1,s_2),\label{e_p_two}
\end{equation}
where $U$ and $V$ are auxiliary random variables with bounded
cardinality.
\end{theorem}

\begin{figure}[h!]{
\psfrag{dots}[][][1.8]{$\cdots$} \psfrag{Dots}[][][2.5]{$\cdots$}
\psfrag{bin}[][][1]{a bin} \psfrag{superbin}[][][1]{a superbin \ }
\psfrag{s2}[][][1]{(contains bins)} \psfrag{b1}[][][1]{(contains
codewords)} \psfrag{t1}[][][1]{$\overbrace{\ \ \ \ \ \ \ \ \ \ \ \ \
}$ } \psfrag{C}[][][1]{a codeword $u^n$ }
\psfrag{t2}[][][0.8]{$2^{n(I(U;S_2)-\epsilon)}$}
\centerline{\includegraphics[width=18cm]{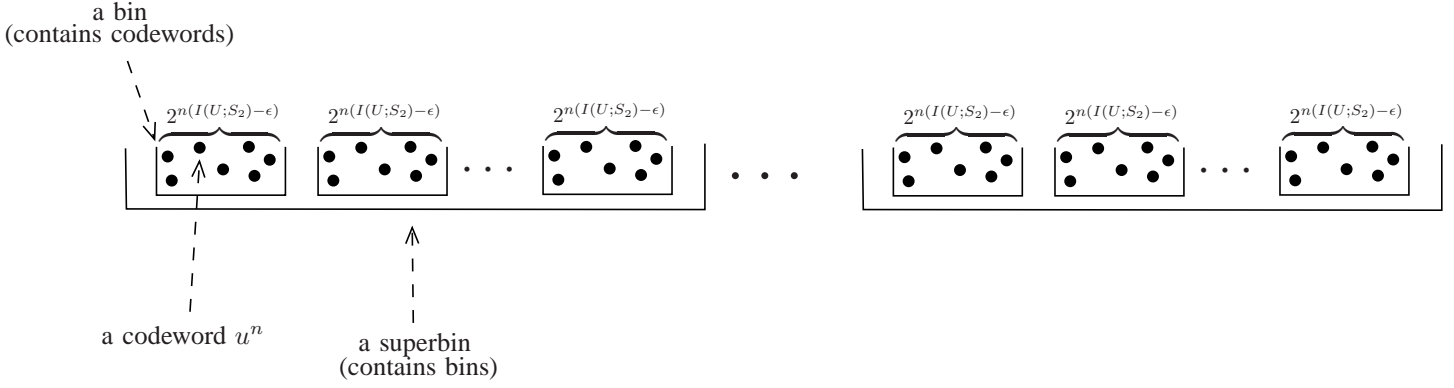}}

\caption{ Double-binning for the achievability of Theorem
\ref{t_two_way_cop}. Double-binning
\cite{LiuMaric_double_binning_08} consists of  two-layer bins, where
in the first layer we have bins that contain codewords and in the
second layer we have superbins that contain bins.}\label{f_superbin}
 }\end{figure}

In the achievability proof of the theorem we use double-binning,
which was introduced by  Liu et al.
\cite{LiuMaric_double_binning_08, LiuPoor_09} to achieve secrecy
capacity in the broadcast channel. Here the double-binning is needed
since one layer of binning will be used for transmitting a common
message between the encoders and an additional layer of binning is
needed for choosing a specific typical sequence using side
information as done in the Wyner-Ziv problem
\cite{Wyner_ziv76_side_info_decoder} and two-way source coding
\cite{Kaspi85_two_way}.  In a double-binning coding scheme we have
special bins that contain other bins rather than codewords, and we
call such a special bin a {\it superbin}, as depicted in Fig.
\ref{f_superbin}.

\begin{proof}


{\bf Achievability part:}

{\it Code construction:} We generate
$2^{n(C_{12}-I(U;S_1)+I(U;S_2)-2\epsilon)}$ superbins, where each
superbin contains $2^{n(I(U;S_1)-I(U;S_2)+2\epsilon)}$ bins, and
each bin contains $2^{n(I(U;S_2)-\epsilon)}$ codewords $U^n$,
generated i.i.d. $\sim P(u)$. Hence, there are
$2^{n(I(U;S_1)+\epsilon)}$ codewords in each superbin and there are
in total $2^{nC_{12}}$ different bins. The index sent from Encoder 1
to Encoder 2 will be a bin number, and the superbin that contains
the bin will represent a common message that is sent from Encoder 1
to Encoder 2.

For each codeword $u^n$, we generate
$2^{n(C_{21}-I(V;S_2|U)+I(V;S_1|U)-2\epsilon)}$ suberbins, where
each superbin contains $2^{n(I(V;S_2|U)-I(V;S_1|U)+2\epsilon)}$
bins, and each bin contains $2^{n(I(V;S_1|U)-\epsilon)}$ codewords
$V^n$, generated i.i.d. $\sim P(v|u)$. Hence, there are
$2^{n(I(V;S_2)+\epsilon)}$ codewords in each superbin and there are
total of $2^{nC_{21}}$ different bins.

For each pair of codewords $(u^n,v^n)$ and for each sequence $s_1^n$
generate $2^{n(R_1-(C_{12}-I(U;S_1)+I(U;S_2)-2\epsilon))}$ codewords
of $X_1^n$ i.i.d. $\sim P(x_1|s_1,u,v)$. Similarly, For each pair of
codewords $(u^n,v^n)$ and for each sequence $s_2^n$ generate
$2^{n(R_2-(C_{21}-I(V;S_2|U)+I(V;S_1|U)-2\epsilon)))}$ codewords of
$X_2^n$ i.i.d. $\sim P(x_2|s_2,u,v)$.

{\it Encoder:} Split  message $m_1\in [1,...,2^{nR_1}]$ into two
messages $m_{1,a}\in
[1,...,2^{n(C_{12}-I(U;S_1)+I(U;S_2)-2\epsilon)}]$ and $m_{1,b}\in
[1,...,2^{n(R_1-(C_{12}-I(U;S_1)+I(U;S_2)-2\epsilon))}].$ 

Associate each message $m_{1,a}$ with a superbin, where in each
superbin there are total of $2^{n(I(U;S_1)+\epsilon)}$  codewords
$u^n$. Search the chosen superbin for a codeword, denoted by
$u^n(m_{1,a},s_1^n)$, with the smallest lexicographical order that
is jointly typical with $s_1^n$ and send its bin number
$[1,...,2^{C_{12}}]$ to Encoder 2. If such a codeword $U^n$ does not
exist, namely, among the codewords in the bin none is jointly
typical with $s_1^n$, choose an arbitrary $u^n$ from the bin (in
such a case the decoder will declare an error). Now, Encoder 2
receives a bin number that contains $2^{n(I(U;S_2)-\epsilon)}$
possible $u^n$ codewords, and looks for the codeword with smallest
lexicographical order that is jointly typical with $s_2^n$. If such
a codeword $U^n$ does not exist, namely, among the codewords in the
bin none is jointly typical with $s_2^n$, choose an arbitrary $u^n$
from the bin (in such a case an error will be declared).

Now, split  message $m_2\in [1,...,2^{nR_2}]$ into two messages
$m_{2,a}\in [1,...,2^{n(C_{12}-I(V;S_2|U)+I(V;S_1|U)-2\epsilon)}]$
and $m_{2,b}\in
[1,...,2^{n(R_2-(C_{21}-I(V;S_2|U)+I(V;S_1|U)-2\epsilon))}]$.

Associate each message $m_{2,a}$ with a superbin, where in each
superbin there are in total $2^{n(I(V;S_2|U)+\epsilon)}$  codewords
$v^n$. Find in the chosen superbin a codeword, denoted by
$v^n(m_{2,a},s_2^n,u^n)$, with the smallest lexicographical order
that is jointly typical with $(s_2^n, u^n)$ and send its bin number
$[1,...,2^{C_{21}}]$ to Encoder 1. If such a codeword $v^n$ does not
exist, namely, among the codewords in the bin none is jointly
typical with $(s_2^n,u^n)$, choose an arbitrary $v^n$ from the bin
(in such a case the decoder will declare an error). Now, Encoder 1
receives a bin number that contains $2^{n(I(V;S_1|U)-\epsilon)}$
possible $v^n$ codewords, and looks for the codeword with the
smallest lexicographical order that is jointly typical with
$(s_1^n,u^n(m_{1,a},s^n))$. If such a codeword $V^n$ does not exist,
namely, among the codewords in the bin none is jointly typical with
$(s_1^n,u^n(m_{1,a},s^n))$, choose an arbitrary $v^n$ from the bin
(in such a case an error will be declared).

Now, Encoder 1 transmits $x_1^n(s_1^n,u^n,v^n,m_{1,b})$, and Encoder
2 transmits $x_2^n(s_2^n,u^n,v^n,m_{2,b})$.

{\it Decoder:} The decoder knows $(s_1^n,s_2^n,y^n)$ and looks for
the indices $\hat m_{1,a}$, $\hat m_{1,b}$, $\hat m_{2,a}$ and $\hat
m_{2,b}$ such that
\begin{equation}
\left(u^n(\hat m_{1,a},s_1^n),v^n(\hat m_{2,a},s_2^n,u^n),
x_1^n(s_1^n,u^n,v^n,\hat m_{1,b}),x_2^n(s_2^n,u^n,v^n,\hat
m_{2,b}),s_1^n,s_2^n,y^n\right)\in
T_{\epsilon}^{(n)}(U,V,X_1,X_2,S_1,S_2,Y).
\end{equation}
If none or more than one such quadruplet is found, an error is
declared. The estimated message sent from Encoder 1 is $(\hat
m_{1,a},\hat m_{1,b})$, and the estimated message transmitted from
Encoder 2 is $(\hat m_{2,a},\hat m_{2,b})$.

{\it Error analysis:} 
Assume $(m_{1,a},m_{2,a},m_{1,b},m_{2,b})=(1,1,1,1)$. Let us define
the event
\begin{equation}
E_{i,j,k,l}\triangleq\left\{ \left(u^n(i,s_1^n),v^n(j,s_2^n,u^n),
x_1^n(s_1^n,u^n,v^n,k),x_2^n(s_2^n,u^n,v^n,l),s_1^n,s_2^n,y^n\right)\in
T_{\epsilon}^{(n)}(U,V,X_1,X_2,S_1,S_2,Y)
\right\}.
\end{equation}

We have an error if either the correct codewords are not jointly
typical with the received sequences, i.e., $E_{1,1,1,1}^c$, or there
exists a different $(i, j,k,l)\neq(1,1,1,1)$ such that $E_{i,j,k,l}$
occurs. From the union of bounds we obtain that
\begin{eqnarray}\label{e_pe2}
P_e^{(n)}&\leq& \Pr(E_{1,1,1,1}^c)+\sum_{i=1,j=1,k=1,l>1}
\Pr(E_{i,j,k,l})+\sum_{i=1,j=1,k>1,l=1}
\Pr(E_{i,j,k,l})+\sum_{i=1,j=1,k>1,l>1}
\Pr(E_{i,j,k,l})\nonumber \\
&& +\sum_{(i,j)\neq 1,k\geq 1,l\geq 1}
\Pr(E_{i,j,k,l}).
\end{eqnarray}
Now let us show that each term in (\ref{e_pe2}) goes to zero as the
blocklength of the code $n$ goes to infinity.
\begin{itemize}
\item Upper-bounding $\Pr(E_{1,1,1,1}^c)$:
Since the total number of codewords in each supperbin associated
with $i$ (or $m_{1,a}$) is larger than $I(U;S_1)$, and since the
codewords were generated i.i.d. $\sim P(u)$, with high probability
there will be at least one codeword that is jointly typical with
$s_1^n$. Let us denote this codeword by $u^n(1,s_1^n).$  Since the
Markov form $U-S_1-S_2$ holds, from the Markov lemma\cite{Berger78}
with high probability $u^n(1,s_1^n)$ would be jointly typical with
$S_2^n$. Furthermore, since each bin in the superbin that is
associated with $i$ contains $2^{n(I(U;S_2)-\epsilon)}$ codewords,
with high probability, there will not be any additional codeword
that is jointly typical with $s_2^n$, hence, Encoder 2 would
identify $u^n(1,s_1^n)$ from the received bin.

Similarly, for a given $u^n\in T^{(n)}_{\epsilon}(U|s_1^n,s_2^n)$,
which is known to Encoder 2, the total number of codewords in each
supperbin associated with $j$ (or $m_{2,a}$) is larger than
$I(V;S_2|U)$, and since the codewords were generated i.i.d.
according to $P(v|u)$, with high probability there will be at least
one codeword that is jointly typical with $(s_2^n,u^n)$. Let us
denote this codeword by $v^n(1,s_2^n,u^n)$ . Since the Markov form
$V-(S_2,U)-S_1$ holds, it follows from the Markov lemma that with
high probability $v^n(1,s_2^n,u^n)$ would be jointly typical with
$(s_1^n,u^n)$. Furthermore, since each bin in the superbin that is
associated with $j$ contains $2^{n(I(V;S_1|U)-\epsilon)}$ codewords,
with high probability, there would not be any additional codeword
that is jointly typical with $(s_1^n,u^n)$, hence, Encoder 2, would
identify $v^n(1,s_2^n,u^n)$ from the bin.

Furthermore, given that $(u^n,v^n,s_1^n,s_2^n)\in T^{(n)}_\epsilon
(U,V,S_1,S_2)$, it follows from the law of large numbers that
$\Pr(E_{1,1,1,1}^c)\to 0$ as $n$ goes to infinity.

\item Upper-bounding $\sum_{i=1,j=1,k=1,l>1}
\Pr(E_{i,j,k,l})$:  The probability that $Y^n$, which is generated
according to $P(y|x_1,s,u,v)$, is jointly typical with $x_2^n$,
which was generated according to
$P(x_2|u,v,s_2)=P(x_2|u,v,s_2,s_1,x_1)$, where
$(x_1^n,s_1^n,s_2^n,u^n,v^n)\in T^{(n)}_\epsilon(X_1,S_1,S_2,U,V)$
is
 upper bounded according to Lemma \ref{l_typical} by
\begin{equation}
\Pr\{(x_1^n,X_2^n,u^n,v^n, s_1^n,s_2^n,Y^n)\in
T^{(n)}_\epsilon|(x_1^n,u^n,v^n,s_1^n,s_2^n)\in
T^{(n)}_\epsilon\}\leq
2^{-n(I(X_2;Y|X_1,S_1,S_2,U,V)-\delta(\epsilon))}.
\end{equation}
Hence, we obtain
\begin{eqnarray}
\sum_{i=1,j=1,k=1,l>1}
\Pr(E_{i,j,k,l})&\leq &2^{n(R_2-(C_{21}-I(V;S_2|U)+I(V;S_1|U)-2\epsilon))}2^{-n(I(X_2;Y|X_1,S_1,S_2,U,V)-\delta(\epsilon))}\nonumber \\
\end{eqnarray}

\item Upper-bounding $\sum_{i=1,j=1,k>1,l=1}
\Pr(E_{i,j,k,l})$: The probability that $Y^n$, which is generated
according to $P(y|x_2,s,u,v)$, is jointly typical with $x_1^n$,
which was generated according to
$P(x_1|u,v,s_1)=P(x_1|u,v,s_2,s_1,x_2)$, where
$(x_2^n,s_1^n,s_2^n,u^n,v^n)\in T^{(n)}_\epsilon(X_2,S_1,S_2,U,V)$
is
 upper bounded according to Lemma \ref{l_typical} by
\begin{equation}
\Pr\{X_1^n,x_2^n,u^n,v^n, s_1^n,s_2^n,Y^n\in
T^{(n)}_\epsilon|x_2^n,u^n,v^n,s_1^n,s_2^n\in T^{(n)}_\epsilon\}\leq
2^{-n(I(X_1;Y|X_2,S_1,S_2,U,V)-\delta(\epsilon))}.
\end{equation}
Hence, we obtain
\begin{eqnarray}
\sum_{i=1,j=1,k>1,l=1}
\Pr(E_{i,j,k,l})&\leq &2^{n(R_1-(C_{12}-I(U;S_1)+I(U;S_2)-2\epsilon))}2^{-n(I(X_1;Y|X_2,S_1,S_2,U,V)-\delta(\epsilon))}\nonumber \\
\end{eqnarray}

\item  Upper-bounding $\sum_{i=1,j=1,k>1,l>1}
\Pr(E_{i,j,k,l})$

\begin{eqnarray}
\lefteqn{\sum_{i=1,j=1,k>1,l>1}
\Pr(E_{i,j,k,l})}\nonumber \\
&\leq
&2^{n(R_1-(C_{12}-I(U;S_1)+I(U;S_2)-2\epsilon)+R_2-(C_{21}-I(V;S_2|U)+I(V;S_1|U)-2\epsilon))}2^{-n(I(X_2,X_1;Y|S_1,S_2,U,V)-\delta(\epsilon))}\nonumber
\\
\end{eqnarray}

\item  Upper-bounding $\sum_{i=1,j=1,k>1,l>1}
\Pr(E_{i,j,k,l})$

\begin{eqnarray}
\sum_{(i,j)\neq 1,k\geq 1,l\geq 1} \Pr(E_{i,j,k,l}) &\leq
&2^{n(R_1+R_2)}2^{-n(I(U,V,X_2,X_1;Y|S_1,S_2)-\delta(\epsilon))}
\end{eqnarray}

\end{itemize}
Finally, we note that if the rate-pair $(R_1,R_2)$ is in the rate
region that is given by (\ref{e_c12_lim_rate})-(\ref{e_p_two}), then
each term in (\ref{e_pe2}) goes to zero as $n\to \infty$; hence
there exists a sequence of codes
$(2^{nR_1},2^{nR_2},2^{nC_{12}},2^{nC_{21}},n)$ such that
$P_{\epsilon}^{(n)}$ goes to zero as $n\to \infty$.


{\bf Converse part:} The converse part combines techniques from
cooperation in a MAC  \cite{Willems83_cooperating} and two-way
source coding \cite{Kaspi85_two_way}. Assume that we have a
$(2^{nR_1},2^{nR_2},2^{nC_{12}},2^{nC_{21}},n)$ code as in
Definition \ref{d_two_way}. We will show the existence of a joint
distribution
$P(s_1,s_2)P(u|s_1)P(v|s_2,u)P(x_1|s_1,u,v)P(x_2|s_2,u,v)P(y|x_1,x_2,s_1,s_2)$
that satisfies~(\ref{e_c12_lim_rate})-(\ref{e_R1+R2_lim_rate})
within some $\epsilon_n$, where  $\epsilon_n$ goes to zero as
$n\to\infty$. Consider
\begin{eqnarray}
nC_{12}&\geq& H(M_{12})\nonumber \\
&=& H(M_{12}|S_2^n)\nonumber \\
&\geq& I(M_{12};S_1^n|S_2^n)\nonumber \\
&\stackrel{}{=}& \sum_{i=1}^n H(S_{1,i}|S_{2,i})-H(S_{1,i}|M_{12},S_1^{i-1},S_{2}^n)\nonumber \\
&\stackrel{(a)}{=}& \sum_{i=1}^n H(S_{1,i}|S_{2,i})-H(S_{1,i}|M_{12},S_1^{i-1},S_{2,i}^n,S_{2,i+1}^n)\nonumber \\
&\stackrel{}{=}& \sum_{i=1}^n I(S_{1,i};M_{12},S_1^{i-1},S_{2,i+1}^n|S_{2,i})\nonumber \\
&\stackrel{(b)}{=}& \sum_{i=1}^n
I(S_{1,i};U_i|S_{2,i}),\label{e_nc12_two}
\end{eqnarray}
 where (a) follows from the Markov chain $S_{1,i}-(M_{12},S_1^{i-1},S_{2,i}^n,S_{2,i+1}^n)-S_{2}^{i-1}$ and (b) follows from the definition
 \begin{equation}\label{e_def_Ui_two_way}
 U_i\triangleq (M_{12},S_1^{i-1},S_{2,i+1}^n).
 \end{equation}
 Now, consider
\begin{eqnarray}
nC_{21}&\geq& H(M_{21})\nonumber \\
&\geq & H(M_{21}|M_{12},S_1^n)\nonumber \\
&\geq& I(M_{21};S_2^n|M_{12},S_1^n)\nonumber \\
&\stackrel{}{=}& \sum_{i=1}^n H(S_{2,i}|S_{2,i+1}^n,S_1^n,M_{12})-H(S_{2,i}|S_{2,i+1}^n,S_1^n,M_{12},M_{21})\nonumber \\
&\stackrel{(a)}{=}& \sum_{i=1}^n H(S_{2,i}|S_{2,i+1}^n,S_1^n,M_{12})-H(S_{2,i}|S_{2,i+1}^n,S_1^{i},M_{12},M_{21})\nonumber \\
&\stackrel{(b)}{=}& \sum_{i=1}^n H(S_{2,i}|U_i,S_{1,i})-H(S_{2,i}|U_i,S_{1,i},V_i)\nonumber \\
&\stackrel{}{=}& \sum_{i=1}^n
I(S_{2,i};V_i|U_i,S_{1,i}),\label{e_nc21_two}
\end{eqnarray}
where (a) follows from the Markov chain
$S_{2,i}-(S_{2,i+1}^n,S_1^{i},M_{12},M_{21})-S_{1,i+1}^n$ and (b)
follows from the definitions of $U_i$ given in
(\ref{e_def_Ui_two_way}) and $V_i$  which is given by
 \begin{equation}\label{e_def_Vi_two_way}
 V_i\triangleq M_{21}.
 \end{equation}
Now, consider
\begin{eqnarray}\label{e_Hm1all_two}
nR_1&=& H(M_{1})\nonumber \\
&=& H(M_{1}|S_1^n,S_2^n,M_2)\nonumber \\
&=& H(M_{1},M_{12}|S_1^n,S_2^n,M_2)\nonumber \\
&=& H(M_{12}|S_1^n,S_2^n,M_2)+H(M_{1}|S_1^n,S_2^n,M_2,M_{12})\nonumber \\
&\leq & H(M_{12}|S_1^n,S_2^n)+H(M_{1}|S_1^n,S_2^n,M_2,M_{12})
\end{eqnarray}
Now, let us consider the terms $H(M_{12}|S_1^n,S_2^n)$ and
$H(M_{1}|S_1^n,S_2^n,M_2,M_{1,2})$ separately.
\begin{eqnarray}\label{e_HM12_two}
H(M_{12}|S_1^n,S_2^n)
&=& H(M_{12}|S_1^n,S_2^n)-H(M_{12})+H(M_{12})\nonumber \\
&\leq & nC_{12}-I(S_1^n,S_2^n;M_{12})\nonumber \\
&\stackrel{(a)}{=}& nC_{12}-I(S_1^n;M_{12}|S_2^n)\nonumber \\
&\stackrel{(b)}{=}& nC_{12}-\sum_{i=1}^n I(S_{1,i};U_i|S_{2,i}),
\end{eqnarray}
where (a) follows from the fact that $M_{12}$ is independent of
$S_2^n$, and (b) follows from (\ref{e_nc12_two}), where it is shown
that $I(S_1^n;M_{12}|S_2^n)=\sum_{i=1}^n I(S_{1,i};U_i|S_{2,i})$.
Now consider the second term,
\begin{eqnarray}\label{e_Hm1_two}
\lefteqn{H(M_{1}|S_1^n,S_2^n,M_2,M_{12})}\nonumber \\
&=& H(M_1|S_1^n,S_2^n,M_2,M_{12},M_{21})\nonumber \\
&\stackrel{(a)}{=}& I(M_1;Y^n|S_1^n,S_2^n,M_2,M_{12},M_{21})+n\epsilon_n \nonumber \\
&=& H(Y^n|S_1^n,S_2^n,M_2,M_{12},M_{21})-H(Y^n|S_1^n,S_2^n,M_2,M_{12},M_{21},M_1)+n\epsilon_n \nonumber \\
&=& H(Y^n|S_1^n,S_2^n,X_2^n,M_2,M_{12},M_{21})-H(Y^n|S_1^n,S_2^n,X_2^n,X_1^n,M_2,M_{12},M_{21},M_1)+n\epsilon_n \nonumber \\
&=& \sum_{i=1}^n H(Y_i|Y^{i-1},S_1^n,S_2^n,X_2^n,M_2,M_{12},M_{21})-H(Y_i|S_1^n,S_2^n,X_2^n,X_1^n,M_2,M_{12},M_{21},M_1,Y^{i-1})+n\epsilon_n \nonumber \\
&\stackrel{(b)}{\leq}& \sum_{i=1}^n H(Y_i|S_{1,i},S_{2,i},X_{2,i},
M_{12},S_1^{i-1},S_{2,i+1}^n,M_{21})-H(Y_i|S_{1,i},S_{2,i},X_{2,i},X_{1,i},
M_{12},S_1^{i-1},S_{2,i+1}^n,M_{21})+n\epsilon_n\nonumber \\
&\stackrel{}{=}& \sum_{i=1}^n
I(X_{1,i};Y_i|S_{1,i},S_{2,i},X_{2,i},U_i,V_i)+n\epsilon_n,
\end{eqnarray}
where (a) follows from Fano's inequality and $\epsilon_n\triangleq
R_1P_e^{(n)}$ and (b) from that fact that conditioning reduces
entropy. Substituting Inequalities (\ref{e_HM12_two}) and
(\ref{e_Hm1_two}) into (\ref{e_Hm1all_two}), we obtain
\begin{eqnarray}
nR_1&\leq& \sum_{i=1}^n
I(X_{1,i};Y_i|S_{1,i},S_{2,i},X_{2,i},U_i,V_i)-I(S_{1,i};U_i|S_{2,i})+nC_{12}.
\end{eqnarray}
Similarly, we obtain
\begin{eqnarray}
nR_2&\leq& \sum_{i=1}^n
I(X_{2,i};Y_i|S_{1,i},S_{2,i},X_{1,i},U_i,V_i)-I(S_{2,i};V_i|S_{1,i},U_i)+nC_{21}.
\end{eqnarray}
Regarding the sum-rate we have
\begin{eqnarray}
nR_1+nR_2&=&H(M_1,M_2)\nonumber \\
&=&H(M_1,M_2|S_1^n,S_2^n)\nonumber \\
&=&I(M_1,M_2;Y^n|S_1^n,S_2^n)+n\epsilon_n \nonumber \\
&\leq&I(X_{1,i},X_{2,i};Y_i|S_{1,i},S_{2,i})+n\epsilon_n,
\end{eqnarray}
and
\begin{eqnarray}
nR_1+nR_2&=&H(M_1,M_2)\nonumber \\
&=&H(M_1,M_2,M_{12},M_{21}|S_1^n,S_2^n) \nonumber \\
&=&H(M_{12}|S_1^n,S_2^n)+H(M_{21}|S_1^n,S_2^n,M_{12})+H(M_1,M_2|S_1^n,S_2^n,M_{12},M_{21})+\epsilon_n,
\end{eqnarray}
and now using (\ref{e_HM12_two}) we bound
\begin{equation}H(M_{12}|S_1^n,S_2^n)\leq nC_{12}-\sum_{i=1}^n I(S_{1,i};U_i|S_{2,i}),\end{equation}
 and similarly
 \begin{equation}H(M_{21}|S_1^n,S_2^n,M_{12})\leq nC_{21}-\sum_{i=1}^n I(S_{2,i};V_i|S_{1,i},U_i).\end{equation}
  Using similar steps as in (\ref{e_Hm1_two}) we bound
  \begin{equation}H(M_1,M_2|S_1^n,S_2^n,M_{12},M_{21})\leq \sum_{i=1}^n
I(X_{2,i},X_{1,i};Y_i|S_{1,i},S_{2,i},U_i,V_i).
\end{equation}
Hence we obtain
\begin{eqnarray*}
nR_1+nR_2&\leq&\sum_{i=1}^n  I(X_{2,i},X_{1,i};Y_i|S_{1,i},S_{2,i},U_i,V_i)-I(S_{1,i};U_i|S_{2,i})-I(S_{2,i};V_i|S_{1,i},U_i)+nC_{12}+nC_{21}.
\end{eqnarray*}
Now we need to verify that the following Markov chains hold:
\begin{equation}(M_{12},S_1^{i-1},S_{2,i+1}^n)-S_{1,i}-S_{2,i},\label{e_m1}\end{equation}
\begin{equation}M_{21}-(S_{2,i},M_{12},S_1^{i-1},S_{2,i+1}^n)-S_{1,i},\label{e_m2}\end{equation}
\begin{equation}X_{1,i}(M_1,S_1^n,M_{21})-(S_{1,i},M_{12},S_1^{i-1},S_{2,i+1}^n,M_{21})-S_{2,i},\label{e_m3}\end{equation}  \begin{equation}X_{2,i}(M_2,S_2^n,M_{12})-(S_{2,i},M_{12},S_1^{i-1},S_{2,i+1}^n,M_{21})-(S_{1,i},X_{1,i}(M_1,S_1^n,M_{21})).\label{e_m4}\end{equation}
Proving the Markov chains (\ref{e_m1})-(\ref{e_m3}) is
straightforward and therefore omitted. To prove the Markov chain in
(\ref{e_m4}), we use the undirected graphical method from
\cite[Section II]{Permuter_steinber_weissman09_two_way_helper}. Fig.
\ref{f_markov_cop} proves the Markov chain
$(M_2,S_2^n)-(S_{2,i},M_{12},S_1^{i-1},S_{2,i+1}^n,M_{21})-(M_1,S_1^n)$,
and as a consequence the Markov chain in (\ref{e_m4}) holds too.
\begin{figure}[h!]{
\psfrag{S11}[][][1]{$S_{1}^{i-1}$} \psfrag{S12}[][][1]{$\ \; S_{1,i}$}
\psfrag{S13}[][][1]{$\ \; S_{1,i+1}^n$}
\psfrag{S21}[][][1]{$\ \ \; S_{2}^{i-1}$} \psfrag{S22}[][][1]{$S_{2,i}\ \ $}
\psfrag{S23}[][][1]{$\ \; S_{2,i+1}^n$}
\psfrag{M1}[][][1]{$\ \  \ \ \ M_1$}
\psfrag{M12}[][][1]{$\ \ \ M_{12}$}
\psfrag{M2}[][][1]{$M_2\ $}
\psfrag{M21}[][][1]{$M_{21}\ \ \ $}

\centerline{\includegraphics[width=9cm]{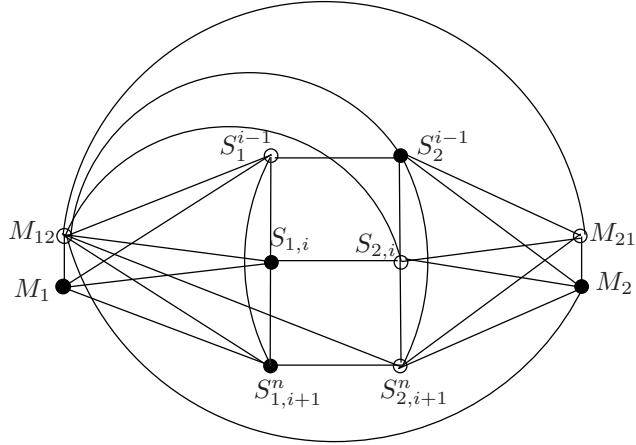}}

\caption{Proof of the Markov chain
$(M_2,S_2^n)-(S_{2,i},M_{12},S_1^{i-1},S_{2,i+1}^n,M_{21})-(M_1,S_1^n)$
using an undirected  graphical technique
\cite{Permuter_steinber_weissman09_two_way_helper}. The undirected
graph corresponds to the joint distribution
$P(s_{1}^{i-1},s_{2}^{i-1})P(s_{1,i},s_{2,i})P(s_{1,i+1}^n,s_{2,i+1}^n)P(m_1)P(m_2)P(m_{12}|m_1,s_1^n)P(m_{21}|m_2,m_{12},s_2^n)$.
The Markov chain follows from the fact that all the paths from
$(M_1,S_1^n)$ to $(M_2,S_2^n)$ go through the nodes
$(S_{2,i},M_{12},S_1^{i-1},S_{2,i+1}^n,M_{21})$.\label{f_markov_cop}
 }}\end{figure}

Finally, let $Q$ be a random variable independent of
$(X^n,S_1^n,S_2^n,Y^n)$, and uniformly distributed over the set
$\{1,2,3,..,n\}$. Define the random variables $U\triangleq(Q,U_Q)$,
$V\triangleq (Q,V_Q)$, and we obtain that the region given by
(\ref{e_c12_lim_rate})-(\ref{e_p_two}) is an outer bound to the set
of all achievable rate-pairs.

To show that the cardinalities of the random variables $U$ and $V$
are bounded we follow similar steps as in Lemma
\ref{l_properties_R_one_way_cop}, first for $U$ and then for $V$. We
note that the cardinality of auxiliary random variables $U$ and $V$
may be bounded by $|\mathcal U|\leq\min(|\mathcal X_1||\mathcal
X_2||\mathcal S_1||\mathcal S_2|+4, |\mathcal Y||\mathcal
S_1||\mathcal S_2|+5), $ and  $|\mathcal V|\leq\min(|\mathcal
X_1||\mathcal X_2||\mathcal S_1||\mathcal S_2||\mathcal U|+3,
|\mathcal Y||\mathcal S_1||\mathcal S_2||\mathcal U|+4).$
\end{proof}

\section{Example and comparison to message-only and state-only cooperation\label{s_example}}

\begin{figure}[h!]{
\psfrag{B}[][][1]{Encoder1} \psfrag{D}[][][1]{Encoder2}
\psfrag{m1}[][][1]{$m_1 \ \ \ \ $}
\psfrag{m2}[][][1]{}
\psfrag{m3}[][][1]{$m_2 \ \ \ \ $}
\psfrag{m4}[][][1]{}
\psfrag{P}[][][1]{$P_{Y|X_1,X_2,S}$} \psfrag{x1}[][][1]{$\ \  X_1$} \psfrag{x2}[][][1]{$\; \ \ \ \ \ \ X_2$} \psfrag{M}[][][1]{BSC($p$)}
\psfrag{s0}[][][1]{$S=0$}
\psfrag{s1}[][][1]{$S=1$}
\psfrag{0}[][][1]{$0$}
\psfrag{1}[][][1]{$1$}
\psfrag{1-p}[][][1]{$1-p$}
\psfrag{s}[][][1]{$S\sim B(\frac{1}{2})$} \psfrag{Yi}[][][1]{$Y$}
\psfrag{W}[][][1]{Decoder} \psfrag{t}[][][1]{$$}

\psfrag{a}[][][1]{a} \psfrag{b}[][][1]{b} \psfrag{c12}[][][1]{$C_{12}$}
\psfrag{c12b}[][][1]{}
\psfrag{Y}[][][1]{$\ \hat m_1$}
\psfrag{Y2}[][][1]{$\hat m_2$}

\centerline{\includegraphics[width=14cm]{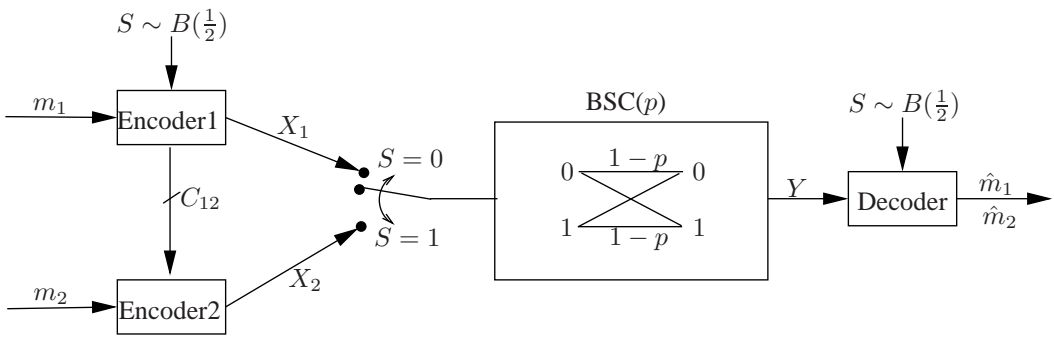}}

\caption{An example of a MAC with one-way cooperation and state
information at one encoder and the decoder. The state $S$ controls
the switch. When $S=0$, $Y=X_1+Z$ and when $S=1$, $Y=X_2+Z$, and
$Z\sim B(p)$.}\label{f_mac_one_way_cop_ex}
 }\end{figure}

Consider the example given in Fig. \ref{f_mac_one_way_cop_ex}, where
the state of the channel controls the switch that determines which
input goes through a binary symmetric channel (BSC) with parameter
$p$. When $S=0$, the binary input $X_1$ goes through and when $S=1$
the binary input $X_2$ goes through, hence the output of the channel
$Y$ is given by
\begin{equation}
Y=\overline S X_1\oplus S X_2\oplus Z,
\end{equation}
where $Z\sim Bernouli (\frac{1}{2})$ and is independent of $S$, the
symbol $\oplus$ denotes XOR, and $\overline S$ denotes $1-S$.
We also have the a constraint on the portion of '1's at the encoders, namely for any pair of codeword $(x_1^n,x_2^n)$,   $\frac{1}{n}\sum_{i=1}^n x_{1,i}\leq p_1$ and $\frac{1}{n}\sum_{i=1}^nx_{2,i}\leq p_2$. Invoking  the following identities
 \begin{eqnarray}
 I(X_1;Y|X_2,S,U)&=&\frac{1}{2}H(X_1\oplus Z|S=0)-\frac{1}{2}H(Z)\nonumber\\
 I(X_2;Y|X_1,S,U)&=&\frac{1}{2}H(X_2\oplus Z|S=1,U)-\frac{1}{2}H(Z)\nonumber\\
 I(X_1,X_2;Y|S,U)&=&\frac{1}{2}H(X_1\oplus Z|S=0)+\frac{1}{2}H(X_2\oplus Z|S=1,U)-H(Z)\nonumber\\
 I(X_1,X_2;Y|S)&=&\frac{1}{2}H(X_1\oplus Z|S=0)+\frac{1}{2}H(X_2\oplus Z|S=1)-H(Z),
 \end{eqnarray}
 we obtain from Theorem \ref{t_mac_one_way_cop} that the capacity region is the set of all rate-pairs $(R_1,R_2)$ that satisfy
\begin{eqnarray}
R_1&\leq & \frac{1}{2}H_b(p_1*p_z)-\frac{1}{2}H_b(p_z)+C_{12}-I(U;S)\nonumber\\
R_2&\leq & \frac{1}{2}H(X_2\oplus Z|S=1,U)-\frac{1}{2}H_b(p_z)\nonumber\\
R_1+R_2&\leq&  \frac{1}{2}H_b(p_1*p_z)+\frac{1}{2}H(X_2\oplus Z|S=1)-H_b(p_z)\label{e_capacity_ex}
\end{eqnarray}
for some conditional distributions $P(u|s)$ and $P(x_2|u)$ where $
I(U;S)\leq C_{12}$. The term $H_b(p)$ denotes the binary entropy
function, which is defined for $0\leq p\leq 1$ as $H_b(p) = -p \log
p - (1 -p) \log(1-p)$.  The term $p*q$ denotes the parameter of a
Bernoulli distribution that results from convolving mod-2 two
Bernoulli distributions with parameters $p$ and $q$, i.e., $p*q =
(1-p)q+(1-q)p$.

\begin{figure}[h!]{
\psfrag{R1}[][][0.8]{$R_1$} \psfrag{R2}[][][0.8]{$R_2$}
\psfrag{0.4}[][][1]{} \psfrag{a}[][][1]{$\uparrow\:$}
\psfrag{c}[][][1]{$\uparrow\:$} \psfrag{g}[][][1]{$\ \ \to$}
\psfrag{e}[][][1]{$\ \ \to$}
\psfrag{f}[][][0.6]{$\frac{1}{2}(H_b(p_2*pz)-H_b(p_z))-C_{12}\ \ \ \
\ \ \ \ \ \ \ \ \ \ \ \ \ \ \ \ \ \ \ \ \ \ \ \ \ \ \ \ \ \ $}
\psfrag{h}[][][0.6]{$\max I(X_2;Y|U,S=1) \ \ \ \ \ \ \ \ \ \ \ \ \ \
\ \ \ \ \ \  \ \ \ \  $}
\psfrag{h2}[][][0.55]{}

\psfrag{b}[][][0.6]{$\ \ \ \ \ \ \ \ \ \ \ \ \ \ \ \ \ \ \ \ \ \ \ \
\ \ \frac{1}{2}(H_b(2p_1*pz)-H_b(p_z))+C_{12}$}
\psfrag{d}[][][0.6]{$\frac{1}{2}(H_b(2p_1*pz)-H_b(p_z))\ \ \ \ \ \ \
\ \ \ \ \ \ \ \ \ \ \ \ \ $}

\centerline{\includegraphics[width=7cm]{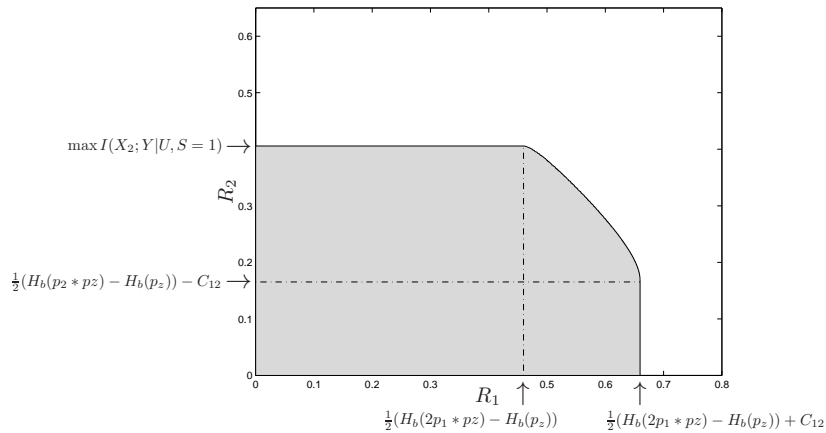}}
\caption{Capacity region of the example depicted in Fig.
\ref{f_mac_one_way_cop_ex} where $C_{12}=0.2$, $p_z=0.01$,
$p1=p2=0.25$.}\label{f_one_capacity_regoion}
 }\end{figure}

Fig. \ref{f_one_capacity_regoion} depicts the capacity region for
the case where $C_{12}=0.2$, $p_z=0.01$ and $p_1=p_2=0.25$. The
capacity region was numerically evaluated using
(\ref{e_capacity_ex}), where the cardinality of the auxiliary random
variable $U$ was assumed to be $|\mathcal U|=2$; changing the
cardinality to 3, 4, or 5 did not increase the numerical capacity
region.

\begin{figure}[h!]{
\psfrag{r1}[][][0.8]{$R_1$}
\psfrag{r2}[][][0.8]{$R_2$}
\psfrag{c0}[][][0.9]{$C_{12}=0 \to \  \ \ \ \ \ \ \ \ \ \ \ \ \ \ \; $}
\psfrag{c0.2}[][][0.9]{$0.2 \to \ \ \ \ \ \ \ \ \ \ \ $}
\psfrag{c0.5}[][][0.9]{$0.5 \to \  \ \ \ \ \ \  \ \ \ \ \ \ \ \ \ \ \ $}
\psfrag{c1}[][][0.9]{$\leftarrow 1 \ \ \ $}

\centerline{\includegraphics[width=9cm]{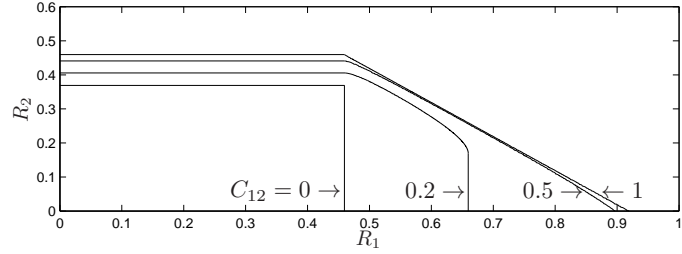}}
\caption{Capacity region of the example depicted in Fig.
\ref{f_mac_one_way_cop_ex} for several values of  $C_{12}$, i.e,
$C_{12}=[0, 0.2,  0.5,  1]$ and $p_z=0.01$,
$p_1=p_2=0.25$.}\label{f_C12_capacity_regoion}
 }\end{figure}

Fig. \ref{f_C12_capacity_regoion} illustrates the influence of the
cooperation rate on the capacity region. It shows the capacity
regions for several rates of cooperation $C_{12}=[0, 0.2,  0.5,  1]$
where $p_z=0.01$, $p_1=p_2=0.25$. One can see that when the
cooperation rate is small an increase in the cooperation rate
significantly influences the capacity region; however, for a large
cooperation rate, such as $C_{12}>0.5$, an increase in the
cooperation rate hardly influences the capacity region.

\begin{figure}[h!]{
\psfrag{B}[][][0.9]{Encoder1} \psfrag{D}[][][0.9]{Encoder2}
\psfrag{E}[][][0.9]{Encoder}
\psfrag{m1}[][][1]{$m_1 \ \ \ \ $}
\psfrag{m2}[][][1]{}
\psfrag{m3}[][][1]{$m_2 \ \ \ \ $}
\psfrag{m4}[][][1]{}
\psfrag{P}[][][1]{$P_{Y|X_1,X_2,S}$} \psfrag{x1}[][][1]{$\ \  X_1$} \psfrag{x2}[][][1]{$\; \ \ \ \ \ \ X_2$} \psfrag{M}[][][1]{BSC($p$)}
\psfrag{s0}[][][1]{$S=0$}
\psfrag{s1}[][][1]{$S=1$}
\psfrag{0}[][][1]{$0$}
\psfrag{1}[][][1]{$1$}
\psfrag{1-p}[][][1]{$1-p$}
\psfrag{s}[][][1]{$S\sim B(\frac{1}{2})$} \psfrag{Yi}[][][1]{$Y$}
\psfrag{S_only}[][][1]{$\ \ \ \ \ \ S$}
\psfrag{W}[][][1]{Decoder} \psfrag{t}[][][1]{$$}

\psfrag{a}[][][1]{a} \psfrag{b}[][][1]{b} \psfrag{c12}[][][1]{$C_{12}$}
\psfrag{c12b}[][][1]{}
\psfrag{Y}[][][1]{$\ \hat m_1$}
\psfrag{Y2}[][][1]{$\hat m_2$}

\centerline{\includegraphics[width=14cm]{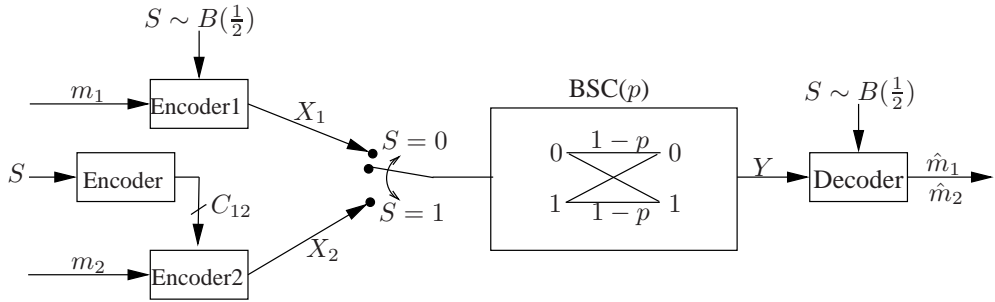}}

\caption{State cooperation. An example inspired by the setting in
\cite{Steinberg_Cemal_MAC05}, where the cooperation is a limited
rate state information and is independent of the message.
}\label{f_mac_ex_state}
 }\end{figure}

\begin{figure}[h!]{
\psfrag{B}[][][0.9]{Encoder1} \psfrag{D}[][][0.9]{Encoder2}
\psfrag{E}[][][0.9]{Encoder}
\psfrag{m1}[][][1]{$m_1 \ \ \ \ $}
\psfrag{m2}[][][1]{}
\psfrag{m3}[][][1]{$m_2 \ \ \ \ $}
\psfrag{m4}[][][1]{}
\psfrag{P}[][][1]{$P_{Y|X_1,X_2,S}$} \psfrag{x1}[][][1]{$\ \  X_1$} \psfrag{x2}[][][1]{$\; \ \ \ \ \ \ X_2$} \psfrag{M}[][][1]{BSC($p$)}
\psfrag{s0}[][][1]{$S=0$}
\psfrag{s1}[][][1]{$S=1$}
\psfrag{0}[][][1]{$0$}
\psfrag{1}[][][1]{$1$}
\psfrag{1-p}[][][1]{$1-p$}
\psfrag{s}[][][1]{$S\sim B(\frac{1}{2})$} \psfrag{Yi}[][][1]{$Y$}
\psfrag{W}[][][1]{Decoder} \psfrag{t}[][][1]{$$}

\psfrag{a}[][][1]{a} \psfrag{b}[][][1]{b} \psfrag{c12}[][][1]{$C_{12}$}
\psfrag{c12b}[][][1]{}
\psfrag{Y}[][][1]{$\ \hat m_1$}
\psfrag{Y2}[][][1]{$\hat m_2$}

\centerline{\includegraphics[width=14cm]{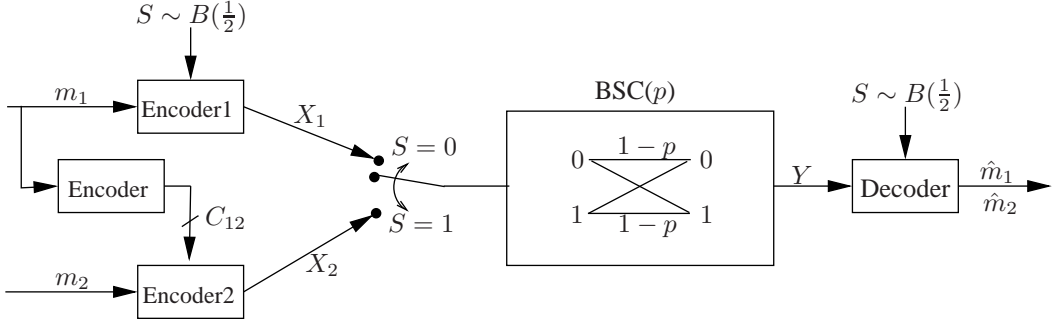}}

\caption{Message cooperation. An example inspired by the setting in
\cite{HaghiAref10_coopertaive_mac_state_fading_isit}, where the
cooperation is a function of the message only, and then after the
cooperation stage the channel state is available to Encoder
1.}\label{f_mac_ex_message}
 }\end{figure}

{\it Comparison to two different kinds of cooperation:} In the
setting analyzed in this paper, we assumed a cooperation link that
may use both the message and the state information. Recent works
assumed similar settings where the cooperation depends only  on the
state \cite{Steinberg_Cemal_MAC05}, as depicted in Fig.
\ref{f_mac_ex_state}, or on the message only
\cite{HaghiAref10_coopertaive_mac_state_fading_isit}
\cite{HaghiAref10_coopertaive_mac_state_fading_IT} as depicted in
Fig. \ref{f_mac_ex_message}.
 For the first case where the cooperation may use only the state information (Fig. \ref{f_mac_ex_state}), the capacity region was derived in \cite{Steinberg_Cemal_MAC05} and may be written as
\begin{eqnarray}
C_{12}&=& I(U;S)  \\
R_1&\leq & I(X_1;Y|X_2,S,U)\\
R_2&\leq & I(X_2;Y|X_1,S,U)  \\
R_1+R_2&\leq& I(X_1,X_2;Y|S,U),\label{e_only_state}
\end{eqnarray}
for some joint distribution of the form
\begin{equation}\label{e_only_state1}
P(s,u)P(x_1|s,u)P(x_2|u)P(y|x_1,x_2,s).
\end{equation}

For the second case where the cooperation may use only the message
(Fig. \ref{f_mac_ex_message}) the capacity region was considered in
\cite{HaghiAref10_coopertaive_mac_state_fading_isit,
HaghiAref10_coopertaive_mac_state_fading_IT} and may be written as
\begin{eqnarray}
R_1&\leq & I(X_1;Y|X_2,S,U)+C_{12}\\
R_2&\leq & I(X_2;Y|X_1,S,U)  \\
R_1+R_2&\leq& \min \left\{ \begin{array}{c}
I(X_1,X_2;Y|S,U)+C_{12},\\ I(X_1,X_2;Y|S) \end{array}
\right\},\label{e_only_message}
\end{eqnarray}
for some joint distribution of the form
\begin{equation}\label{e_only_message2}
P(s)P(u)P(x_1|s,u)P(x_2|u)P(y|x_1,x_2,s),
\end{equation}
where $U$ and $V$ are auxiliary random variables with bounded
cardinality.

Both regions, the one in (\ref{e_only_state})-(\ref{e_only_state1})
and the one in (\ref{e_only_message})-(\ref{e_only_message2}), are
contained in the region of Theorem \ref{t_mac_one_way_cop} where the
cooperation may use both the message and the state. It is
interesting to note that one can obtain the regions
(\ref{e_only_state})-(\ref{e_only_state1}), and
(\ref{e_only_message})-(\ref{e_only_message2}) by adding only an
additional constraint  to the region of Theorem
\ref{t_mac_one_way_cop}. More precisely,  to obtain the regions
(\ref{e_only_state})-(\ref{e_only_state1}) add the constraint
$C_{12}= I(U;S)$, and to obtain the region
(\ref{e_only_message})-(\ref{e_only_message2}) add the constraint
$I(U;S)=0$ to the region (\ref{e_c1})-(\ref{e_c5}) of Theorem
\ref{t_mac_one_way_cop}.

\begin{figure}[h!]{
\psfrag{r1}[][][0.9]{$\ \  R_1$} \psfrag{r2}[][][0.9]{$R_2$}
\psfrag{0.4}[][][1]{} \psfrag{0.2}[][][1]{}\psfrag{0.6}[][][1]{}
\psfrag{0.9}[][][1]{}\psfrag{0.8}[][][1]{}
\psfrag{a}[][][1]{$\uparrow\:$}\psfrag{i}[][][1]{$\uparrow\:$}
\psfrag{c}[][][1]{$\uparrow\:$} \psfrag{g}[][][1]{$\ \ \to$}
\psfrag{e}[][][1]{$\ \ \to$} \psfrag{cs}[][][1]{ \color{blue} State
coop. $\to \ \ \ \ \ \ \ \ \ \ \ \ \ \ \ \ \ \ \ \  $}
\psfrag{cm}[][][1]{\color{green} Message coop. $\to\ \ \ \ \ \ \ \ \
\ \  \ \ \ \ \ \ \ \ \ \ \ \ \ \ \ \ \ \ \ \ \ \ \ \ \ \ $}
\psfrag{csm}[][][1]{\color{red} $ \ \ \ \ \ \ \ \ \ \ \ \ \ \ \ \ \
\ \ \ \ \leftarrow $Message\&state coop. }

\psfrag{f}[][][0.6]{$\frac{1}{2}(H_b(p_2)-H_b(p_z)) \ \ \ \ \ \ \ \ \ \ \ \ \ \ \ \ \ \ \ \ \ $}
\psfrag{h}[][][0.6]{$\max I(X_2;Y|U,S=1) \ \ \ \ \ \ \ \ \ \ \ \ \ \
\ \ \ \ \ \  \ \ \ \  $}

\psfrag{b}[][][0.6]{$\ \ \ \ \ \ \ \ \ \ \ \ \ \ \ \ \ \ \ \ \ \ \ \
\ \ \ \max I(X_1,X_2;Y|S)$}
\psfrag{j}[][][0.6]{$\frac{1}{2}(H_b(2p_1)-H_b(p_z))\ \ \ \ \ \ \ \
\ \ \ \ \ \ \ \ \ \ \ \ \ \ $}
\psfrag{d}[][][0.6]{$\frac{1}{2}(H_b(2p_1)-H_b(p_z)+H_b(p_2)-H_b(p_z))\
\  \ \ \ \ \ \ \ \ \ \ \ \ \ \ \ \ \ \ \ \ \ \ \
\ \ \ \ \ \ \ \ \ \ \ \ $}

\centerline{\includegraphics[width=12cm]{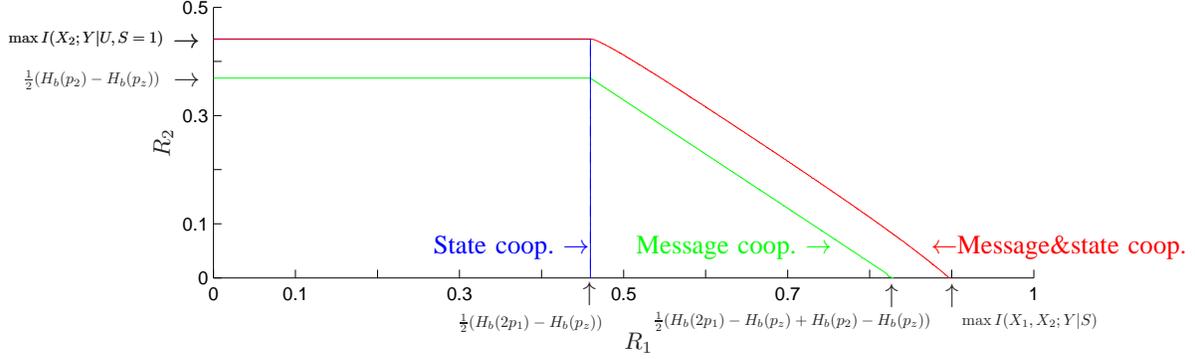}}
\caption{The regions of three settings with a cooperation link
$C_{12}=0.5$. The blue region corresponds to the case where the
cooperation is based only on the state information as depicted in
Fig. \ref{f_mac_ex_state}. The green region corresponds to the case
where the cooperation is based only on the message and not on the
state as depicted in Fig. \ref{f_mac_ex_message}. Finally, the red
region is the one that corresponds to the setting of this paper
where the cooperation may use both the state and the message as
depicted in Fig. \ref{f_mac_one_way_cop_ex}.
}\label{f_three_capacity_regoion}
 }\end{figure}

Fig. \ref{f_three_capacity_regoion} depicts the capacity regions
obtained for a cooperation link $C_{12}=0.5$ for the three settings:
\begin{enumerate}
\item state-cooperation, where the cooperation is based only on the
state information  (Fig. \ref{f_mac_ex_state}),
\item message-cooperation, where the cooperation is based only on the
message (Fig. \ref{f_mac_ex_message}),
\item  message-state
cooperation, where the cooperation may use both the state and the
message (Fig. \ref{f_mac_one_way_cop_ex}).
\end{enumerate}
In this example, one can note from  Fig.
\ref{f_three_capacity_regoion}  that state-cooperation increases the
capacity region only in the direction of $R_2$, message-cooperation
increases the capacity region only in the direction of $R_1$, and
message-state cooperation increases the capacity region  in the
direction of both $R_1$ and $R_2$.

From the comparison above, it is interesting to note that there are
special cases where the state-only cooperation or the message-only
cooperation performs as well as the
combined state-message cooperation. 

{\it Equal rates, i.e.,  $R_1=R_2$:} consider the example of one-way
cooperation depicted in Fig. \ref{f_mac_one_way_cop_ex}, where we
are interested in equal-rates working-point, i.e., $R_1=R_2$. Since
on the boundary region $R_2\leq R_1$, the best equal-rate working
point is achieved by maximizing $R_2$. To maximize $R_2$ in one-way
cooperation, there is no need for message cooperation and therefore
the state-only cooperation  achieves the maximum equal rate point.

{\it Effectively, no power constraint,  $p_1=p_2=0.5$:} consider the
one-way cooperation as depicted in Fig. \ref{f_mac_one_way_cop_ex},
where, effectively, there is no power constraint; this means that
$p_1,p_2$ may be equal to or larger than $0.5$. For this case, the
state information at the transmitter does not enlarge the rate
region, hence the message-only cooperation as introduced by Willems
\cite{Willems83_cooperating} is optimal.

\subsection{Splitting the cooperation link in  message-only and and state-only links}
\begin{figure}[h!]{
\psfrag{B}[][][0.9]{Encoder1} \psfrag{D}[][][0.9]{Encoder2}
\psfrag{E}[][][0.9]{Encoder} \psfrag{m1}[][][1]{$m_1 \ \ \ \ $}
\psfrag{m2}[][][1]{}
\psfrag{m3}[][][1]{$m_2 \ \ \ \ $}
\psfrag{m4}[][][1]{}
\psfrag{P}[][][1]{$P_{Y|X_1,X_2,S}$} \psfrag{x1}[][][1]{$\ \  X_1$}
\psfrag{x2}[][][1]{$\; \ \ \ \ \ \ X_2$} \psfrag{M}[][][1]{BSC($p$)}
\psfrag{s0}[][][1]{$S=0$} \psfrag{s1}[][][1]{$S=1$}
\psfrag{0}[][][1]{$0$} \psfrag{1}[][][1]{$1$}
\psfrag{1-p}[][][1]{$1-p$} \psfrag{s}[][][1]{$S\sim B(\frac{1}{2})$}
\psfrag{Yi}[][][1]{$Y$} \psfrag{W}[][][1]{Decoder}
\psfrag{t}[][][1]{$$}

\psfrag{a}[][][1]{a} \psfrag{b}[][][1]{b}
\psfrag{c12m}[][][1]{$C_{12}^m\ \ $} \psfrag{c21m}[][][1]{$C_{21}^m\
 \ $}

\psfrag{c12s}[][][1]{$C_{12}^s$} \psfrag{S_only}[][][1]{$\ \ \ \ S$}

\psfrag{c12b}[][][1]{}
\psfrag{Y}[][][1]{$\ \hat m_1$} \psfrag{Y2}[][][1]{$\hat m_2$}

\centerline{\includegraphics[width=14cm]{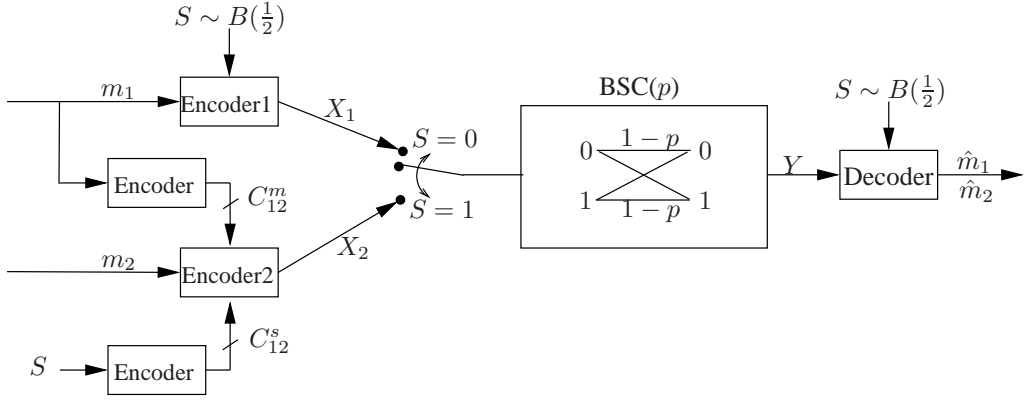}}


\caption{Separate message and state cooperation. There are two
cooperation links at rates $C_{12}^m$ and  $C_{12}^s$. The first
link uses only the message information $m_1$, the second link uses
only the state information $S^n$.}\label{f_mac_sep_s_m}
 }\end{figure}

In this subsection we investigate what happens if we split the
cooperation link into two links: one link for message-only
cooperation at rate $C_{12}^m$ and the other link for state-only
cooperation at rate $C_{12}^s$ as shown in Fig. \ref{f_mac_sep_s_m}.
We derive the capacity region for this setting and show that the
split is strictly suboptimal.

\begin{theorem}\label{t_mac_s_m_sep}
The capacity region of the MAC  with separated links, for message
and state cooperation, as shown in Fig. \ref{f_mac_sep_s_m} is the
closure of the set that contains all rates that satisfy
\begin{eqnarray}
C_{12}^s&\geq& I(U;S) \label{e_c1_s} \\
R_1&\leq & I(X_1;Y|X_2,S,U,V)+C_{12}^m\\
R_2&\leq & I(X_2;Y|X_1,S,U,V) \\
R_1+R_2&\leq& \min \left\{ \begin{array}{c}
I(X_1,X_2;Y|S,U,V)+C_{12}^m,\\ I(X_1,X_2;Y|S,U) \end{array}
\right\},\label{e_R1+R2_s}
\end{eqnarray}
for some joint distribution of the form
\begin{equation}\label{e_c5_s}
P(s,u)P(v)P(x_1|s,u,v)P(x_2|u,v)P(y|x_1,x_2,s).
\end{equation}

\end{theorem}

The proof of the theorem for the case where the MAC is of the
general form $P(y|x_1,x_2,s)$ is given in the appendix. The converse
is based on the identification of the auxiliary random variable $U$
being a function of the state sequence only, and the identification
of the auxiliary random variable $V$ being a function of the message
$M_1$ only; hence the pair $(U,S^n)$ is independent of $V$, since
$M_1$ is independent of $S^n$. The achievability part is based on
generating the coordination $P_{S,U}$ and then multiplexing the
cooperation MAC codebooks according to $U$. There is no need for
binning in the achievability part where the coopertaion link is
split.

\begin{figure}[h!]{
\psfrag{r1}[][][0.9]{$\ \  R_1$} \psfrag{r2}[][][0.9]{$R_2$}
\psfrag{0.4}[][][1]{} \psfrag{0.2}[][][1]{}\psfrag{0.6}[][][1]{}
\psfrag{0.9}[][][1]{}\psfrag{0.7}[][][1]{}\psfrag{0.8}[][][1]{}
\psfrag{a}[][][1]{$\uparrow\:$}\psfrag{i}[][][1]{$\uparrow\:$}
\psfrag{c}[][][1]{$\uparrow\:$} \psfrag{g}[][][1]{$\ \ \to$}
\psfrag{e}[][][1]{$\ \ \to$} \psfrag{cs0.25}[][][1]{ \color{blue}
separate state and message coop. $\to   $\ \ \ \ \ \ \ \ \ \ \ \ \ \
\ \ \  \ \ \ \ \ \ \ \ \ \ \ \ \  \ \ \ \ \ \  \ \ \ \ \  \ \ \ \ \
\  \ \ \ \  \ \ \ \ \ \ \ \ \ \ \ \ \ \  \ \ }
\psfrag{csm}[][][1]{\color{red} $ \leftarrow $Message\&state coop. \
\ \ \ \ \ \ \ \ \ \ \ \ } \psfrag{t1}[][][1]{\color{red} \ \ \ \ \ \
\ \ \ \ \ \ \ \ \ \ \ \ \ \ \ \ \  $C_{12}=0.5$}

\psfrag{t2}[][][1]{\color{blue}  $C_{12}^s=0.25$ $C_{12}^m=0.25$\ \
\ \ \ \ \ \ \ \ \ \ \ \ \ \ \ \ \ \ \ \ \ \ \ \ \ \ \  \ \ \ \ \ \ \
\ \ \ \ \ \ \ \ \ \  \ \ \ \ \ \ \ \ \ \ \ \ \ \ \ \ \ \ \ \  \ \ \
\ \ \  \ \ \ \  \ \ \ \ \  \ \ \ \ \  \ \ \ \ \ \  \ \ }

\psfrag{h}[][][0.6]{$\max I(X_2;Y|U,S=1) \text{\  s.t. \ }
I(U;S)\leq 0.5 \ \ \ \ \ \ \ \ \ \ \ \ \ \ \ \ \ \ \ \  \ \ \ \ \ \
\ \ \ \ \ \ \ \ \ \ \ \ \ \ \ \ \ \  \  $} \psfrag{f}[][][0.6]{$\max
I(X_2;Y|U,S=1) \text{\  s.t. \ } I(U;S)\leq 0.25 \ \ \ \ \ \ \ \ \ \
\ \ \ \ \ \ \ \ \ \  \ \ \ \ \ \ \ \ \ \ \ \ \ \ \ \ \ \ \ \ \ \ \ \
\ $}

\psfrag{b}[][][0.6]{$ \ \ \ \ \ \ \ \ \ \ \ \ \ \ \ \ \max
I(X_1,X_2;Y|S)$}
\psfrag{j}[][][0.6]{$\frac{1}{2}(H_b(2p_1)-H_b(p_z))\ \ \ \ \ \ \ \
\ \ \ \ \ \ \ \ \ \ \ \ \ \ $}
\psfrag{l}[][][0.6]{$\frac{1}{2}(H_b(2p_1)-H_b(p_z))+0.25 \ \ \ \ \
\ \ \  \ \ \ \ $}

\psfrag{d}[][][0.6]{$\frac{1}{2}(H_b(2p_1)-H_b(p_z)+H_b(p_2)-H_b(p_z))\
\  \ \ \ \ \ \ \ \ \ \ \ \ \ \ \ \ \ \ \ \ \ \ \
\ \ \ \ \ \ \ \ \ \ \ \ $}

\centerline{\includegraphics[width=12cm]{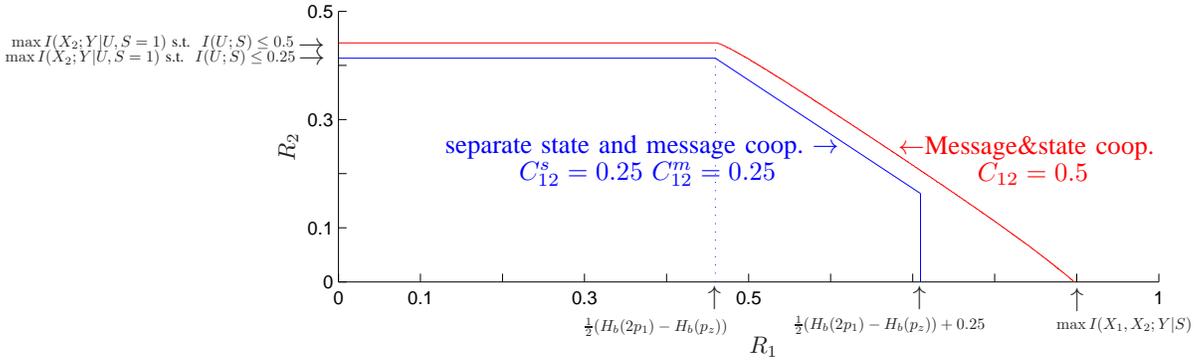}}
\caption{Comparison between the capacity regions of separate state
and message cooperation and joint state-message cooperation. The
blue line corresponds to the capacity region of the setting in Fig.
\ref{f_m_s_seperate_regoion} where $C_{12}^s=C_{12}^m=0.25.$ The red
line corresponds to the capacity region of the setting in Fig.
\ref{f_mac_ex_message} where $C_{12}=0.5$.
}\label{f_m_s_seperate_regoion}
 }\end{figure}

Fig. \ref{f_m_s_seperate_regoion} depicts the capacity region of the
example where separate state cooperation and message cooperation
exists (i.e., the setting of Fig. \ref{f_mac_sep_s_m}) and
$C_{12}^s=C_{12}^m=0.25$. From Fig. \ref{f_m_s_seperate_regoion} we
learn that using the naive strategy of splitting the cooperation
link into message-only cooperation and state-only cooperation is
strictly suboptimal.

\bibliographystyle{unsrt}
\bibliographystyle{IEEEtran}

\appendix

Here we present the proof of Theorem \ref{t_mac_s_m_sep}, where  MAC
is of the general form $P(y|x_1,x_2,s)$.

{\it Proof of Theorem \ref{t_mac_s_m_sep}:}

{\bf Converse part:} Assume that we have a
$(2^{nR_1},2^{nR_2},2^{nC_{12}^m},2^{nC_{12}^s},n)$ code. We will
show the existence of a joint distribution
$P(s)P(u|s)P(v)P(x_1|s,u,v)P(x_2|u,v)P(y|x_1,x_2,s)$ that
satisfies~(\ref{e_c1_s})-(\ref{e_R1+R2_s}) within some $\epsilon_n$,
where  $\epsilon_n$ goes to zero as $n\to\infty$. Let $M_{12}^s\in
\{1,2,...,2^{nC_{12}^s}\}$ and $M_{12}^m\in
\{1,2,...,2^{nC_{12}^m}\}$ be the message sent on the state
cooperation link and the message cooperation link, respectively.
Consider
\begin{eqnarray}
nC_{12}^s&\geq& H(M_{12}^s)\nonumber \\
&\geq& I(M_{12}^s;S^n)\nonumber \\
&\stackrel{(a)}{=}& \sum_{i=1}^n I(S_i;M_{12}^s,S^{i-1})\nonumber \\
&\stackrel{(b)}{=}& \sum_{i=1}^n I(S_i;U_i),\label{e_nc12_one}
\end{eqnarray}
where (a) follows from the fact that $S_i$ is i.i.d. and (b) follows
from the definition of $U_i$, which is
\begin{equation}\label{e_def_Ui}
U_i \triangleq (M_{12}^s,S^{i-1}).
\end{equation}
Now, consider
\begin{eqnarray}\label{e_Hm1all}
nR_1&=& H(M_{1})\nonumber \\
&\stackrel{(a)}{=}& H(M_{1},M_{12}^m)\nonumber \\
&=& H(M_{12}^m)+H(M_{1}|M_{12}^m)\nonumber \\
&\stackrel{(b)}{\leq}& nC_{12}^m+H(M_{1}|M_{12}^m,M_2,S^n,M_{12}^s)\nonumber \\
&\stackrel{(c)}{\leq}& nC_{12}^m+I(M_{1};Y^n|M_{12}^m,M_2,S^n,M_{12}^s)+n\epsilon_n\nonumber \\
&=& nC_{12}^m+I(M_{1},X_1^n;Y^n|M_{12}^m,M_2,S^n,X_2^n)+n\epsilon_n\nonumber \\
&\leq& nC_{12}^m+\sum_{i=1}^n I(X_{1,i};Y_i|M_{12}^m,M_{12}^s,S^i,X_{2,i})+n\epsilon_n\nonumber \\
&\stackrel{(d)}{=}&nC_{12}^m+\sum_{i=1}^n
I(X_{1,i};Y_i|V_i,U_i,S_i,X_{2,i})+n\epsilon_n,
\end{eqnarray}
where (a) follows from the fact that $M_{12}^m$ is a deterministic
function of $M_1$, (b) from the fact that $S^n$ is independent of
$M_1$ and $M_{12}^s$ is a deterministic function of $S^n$, (c) from
Fano's inequality and the definition of $\epsilon_n\triangleq
(R_1+R_2)P_e^{(n)}$. Step (d) follows from the definition of the
auxiliary random variable
\begin{equation}
V_i\triangleq M_{12}^m.
\end{equation}
Now using similar steps as above we obtain the following additional
upper bounds
\begin{eqnarray}\label{e_Hm1all}
nR_2&=& H(M_{2})\nonumber \\
&=& H(M_2|M_1,M_{12}^m,S^n,M_{12}^s)\nonumber \\
&\leq& I(M_2;Y^n|M_1,M_{12}^m,S^n,M_{12}^s)+n\epsilon_n\nonumber \\
&\leq&\sum_{i=1}^n I(X_{2,i};Y_i|X_{1,i},V_i,U_i,S_i)+n\epsilon_n \nonumber \\
\end{eqnarray}
and
\begin{eqnarray}
nR_1+nR_2&=&H(M_1,M_2,M_{12}^m)\nonumber \\
&=&H(M_1,M_2,M_{12}^m|S^n)\nonumber \\
&=&H(M_{12}^m|S^n)+H(M_1,M_2|M_{12}^m,S^n,M_{12}^s)\nonumber \\
&\leq&nC_{12}^m+I(M_1,M_2;Y^n|M_{12}^m,S^n,M_{12}^s)+n\epsilon_n\nonumber \\
&\leq& nC_{12}^m+\sum_{i=1}^n
I(X_{1,i},X_{2,i};Y_i|V_i,U_i,S_i)+n\epsilon_n ,
\end{eqnarray}
and
\begin{eqnarray}
nR_1+nR_2&=&H(M_1,M_2,M_{12}^m)\nonumber \\
&=&H(M_1,M_2,M_{12}^m|S^n)\nonumber \\
&\leq&I(M_1,M_2;Y^n|S^n,M_{12}^s)+n\epsilon_n\nonumber \\
&\leq& \sum_{i=1}^n I(X_{1,i},X_{2,i};Y_i|U_i,S_i)+n\epsilon_n \nonumber \\
\end{eqnarray}
Now,  we note that $V_i$ is independent of $(U_i,S_i)$ since $M_1$
is independent of $S^n$, and $X_2-(U,V)-(X_1,S_1)$ is a Markov chain
since
$(M_2,M_{12}^m,M_{12}^s)-(S^{i-1},M_{12}^m,M_{12}^s)-(M_1,S^n)$
holds. Finally, let $Q$ be a random variable independent of
$(X^n,S^n,Y^n)$, and uniformly distributed over the set
$\{1,2,3,..,n\}$. Define the random variables $U\triangleq(Q,U_Q)$,
$V\triangleq (Q,V_Q)$, and we obtain that the region given by
(\ref{e_c1_s})-(\ref{e_c5_s}) is an outer bound to the achievable
region.

To show that the cardinality of the random variables $U$ and $V$ is
bounded we follow similar steps as in Lemma
\ref{l_properties_R_one_way_cop}, first for $U$ and then for $V$. We
note that the cardinality of auxiliary random variables $U$ may be
bounded by $|\mathcal U|\leq|\mathcal S|+4$ and for auxiliary random
variables $V$, we have $|\mathcal V|\leq\min(|\mathcal X_1||\mathcal
X_2||\mathcal S||\mathcal U|+3, |\mathcal Y||\mathcal S||\mathcal
U|+4).$

{\bf Outline of achievability part:} The achievability part is
straightforward once we observe that we can generate a coordination
$P_{U,S}$ with a rate $C_{12}^s>I(U;S)$ and then use a multiplexer
where $U^n$, which is known to all encoders and to the decoder, is
the control sequence of the multiplexer. For a given $U=u$ we obtain
a MAC with cooperation $C_{12}^m$ where the state is known to one
encoder and to the decoder, hence the region
\begin{eqnarray}
R_1&\leq & I(X_1;Y|X_2,S,U=u,V)+C_{12}^m\\
R_2&\leq & I(X_2;Y|X_1,S,U=u,V) \\
R_1+R_2&\leq& \min \left\{ \begin{array}{c}
I(X_1,X_2;Y|S,U=u,V)+C_{12}^m\\ I(X_1,X_2;Y|S,U=u) \end{array}
\right\},\label{e_R1+R2_s}
\end{eqnarray}
is achievable. Averaging over $P(u)$ results in the region given by
(\ref{e_c1_s})-(\ref{e_c5_s}).
 \hfill\QED

\end{document}